\documentclass[prb,twocolumn,aps,longbibliography, superscriptaddress, nofootinbib]{revtex4-1}

\usepackage[dvipsnames]{xcolor}

\usepackage{graphicx}
\usepackage{ntheorem}
\usepackage{mathrsfs}

\usepackage{hyperref}
\usepackage{comment}
\usepackage{bbm}

\usepackage{amsmath}
\usepackage{amssymb}
\usepackage{longtable} 
\usepackage{boldline}
\usepackage{tikz}
\usetikzlibrary{calc,shapes}

\usepackage{placeins}

\newcommand{\defl}{\mathrel{\mathop:}=}
\newcommand{\defr}{\mathrel={\mathop:}}
\newcommand{\rddots}{\reflectbox{$\ddots$}}
\usepackage{array}

\definecolor{indigo}{rgb}{0.29, 0.0, 0.51}
\definecolor{RGBgreen}{RGB}{ 18 173 42}

\newcommand{\ycal}{{\scriptstyle{\mathcal{Y}}}}

\begin{document}

\author{Nico Leumer}
\author{Milena Grifoni}
\affiliation{Institute for Theoretical Physics, University of Regensburg, 93053 Regensburg, Germany}
\author{Bhaskaran Muralidharan}
\affiliation{Department of Electrical Engineering, Indian Institute of Technology Bombay, Powai, Mumbai-400076, India}
\author{Magdalena Marganska}
\affiliation{Institute for Theoretical Physics, University of Regensburg, 93053 Regensburg, Germany}
 
\title{Linear and non-linear transport across a finite Kitaev chain: an exact analytical study}
\begin{abstract}
We present exact analytical results for the differential conductance of a finite Kitaev chain in an N-S-N configuration, where the topological superconductor is contacted on both sides with normal leads. Our results are obtained with the Keldysh non-equilibrium Green's functions technique, using the full spectrum of the Kitaev chain without resorting to minimal models. 
A closed formula for the linear conductance is given, and the analytical procedure to obtain the differential conductance for the transport mediated by higher excitations is described. 
The linear conductance attains the maximum value of $e^2/h$ only for the exact zero energy states. Also the differential conductance exhibits a complex pattern created by numerous crossings and anticrossings in the excitation spectrum. We reveal the crossings to be protected by inversion symmetry, while the anticrossings result from a pairing-induced hybridization of particle-like and hole-like solutions with the same inversion character.
Our comprehensive treatment of the Kitaev chain allows us also to identify the contributions of both local and non-local transmission processes to transport at arbitrary bias voltage. Local Andreev reflection processes dominate the transport within the bulk gap and diminish for higher excited states, but reemerge when the bias voltage probes the avoided crossings. The non-local direct transmission is enhanced above the bulk gap, but contributes also to the transport mediated by the topological states.
\end{abstract}
 
\maketitle

\section{Introduction}

The search for Majorana zero modes (MZM) in topological superconductor systems is currently an intensely pursued quest in condensed matter physics,\cite{Alicea-2010,Leijnse-2012,Elliott-2015,Aguado} with the primary aim to realize a robust framework for topological quantum computing.\cite{Das-Sarma-2015,Aasen-2016,obrien:prl2018} Currently the most advanced experimentally platform for Majorana devices are based on proximitized semiconducting nanowires\cite{Deng-2016,Gul-2018,Prada-2020}, although they have not yet been unambiguously proven to host Majorana states.
Transport properties of Majorana nanowire devices have been intensively studied, with the main purpose of devising a detection scheme for the Majorana states by determining their transport fingerprints.
The most fundamental one, that of observing a quantized zero bias peak in conductance,\cite{Law-2009,Flensberg-2010,Aguado} can be mimicked by trivial Andreev bound states\cite{Liu-2017,Prada-2020,Vuik-2019,Pan-2020} or level repulsion in multiband systems,\cite{Chen-2019,Woods-2019}, thus several detection schemes exploiting also Majorana non-locality have been proposed.\cite{Deng-2018,Hell-2018,Zhang-2019} From the point of view of the applications, one of the schemes for the readout of Majorana qubits is based on transport interferometry,\cite{Plugge-2017,Qin-2019} providing further motivation to explore the transport properties of Majorana devices.

Most of the works in this domain are out of necessity either numerical or based upon a minimal model, concentrating on charge transport through the in-gap states.\cite{Liu-2011,Lim-2012,Rainis-2013,Prada-2017,Liu-2017,Prada-2020,Vuik-2019,Pan-2020} Our aim is to find an analytical expression for the current flowing through a topological superconductor, taking into account its full excitation spectrum. The knowledge of such analytical solutions for at least one topological superconductor is instrumental in testing the reliability of the numerical results. As our model system we take a prototypical topological superconductor, the Kitaev chain.\cite{kitaev:physusp2001} 
Although the low energy spectrum of a Kitaev chain with two Majorana states has served as the basis for the minimal models of nanowire transport, we are aware of only few analytical studies which focused on the transport characteristics of the Kitaev chain itself, achieving its description in analytical terms for several parameter ranges.\cite{Stoof-Kitaev,Giuliano-2018} 
Doornenbal et al.\cite{Stoof-Kitaev} treat the chain as a fragment of an N-S-N system, but with the bias drop occurring at one contact only, which yields the well known value $2e^2/h$ for the conductance through an MZM. Without a self-consistent calculation it leads however also to non-conservation of current.
Another recent work\cite{Giuliano-2018} studied the low energy transport properties of a Kitaev chain with long-range superconducting pairing, using a Green's functions technique combined with the scattering matrix approach. The transport calculation is analytical, although it needs as input the eigenvectors, which are obtained from a numerical diagonalization of the Hamiltonian. 

In this work we use the Keldysh non-equilibrium Green's functions technique (NEGF) and the notion of Tetranacci polynomials to derive
analytical expressions for both the current and conductance of a Kitaev chain in an N-S-N configuration, in the linear as well as non-linear transport regime, for arbitrary hopping $t$, superconducting pairing $\Delta$ and chemical potential $\mu$. Thus we can access not only the known transport properties of the topological states, but also of the higher excited states.  

While we derive the differential conductance for arbitrary bias drop at the contacts, we show only the results for symmetric bias -- the configuration in which the current is conserved. In consequence, crossed Andreev reflection processes do not contribute to transport. For the chosen symmetric setup, the transport occurs via two mechanisms, local Andreev reflection and non-local direct transmission. Both contributions feature conductance peaks resonant with the excitation energies, but with different weights.
The transport within the bulk gap is dominated by the local, Andreev processes, while the main contribution to transport above the bulk gap comes from the non-local, direct transmission.
The excitation spectrum above the bulk gap contains several series of crossings and avoided crossings, ubiquitous in the spectra of Majorana nanowires,\cite{Liu-2017,Pan-2020,Chen-2019,Prada-2017,Mishmash-2016,Kobialka-2019,Danon-prl2020}. Our analysis sheds light onto the nature of these states features. We find that the crossings are protected by the inversion symmetry of the normal chain -- the degenerate eigenstates have particle (hole) sectors of opposite inversion character. On the other hand, the particle (hole) sectors of {\em anti}-crossing states match under inversion, and the superconducting pairing allows the particle-like and hole-like solutions of the linear chain to hybridize.
Inside the anticrossings the Andreev reflection processes are revived, reminiscent of their importance in the subgap transport. Similarly, even though the direct transmission plays the prominent role in the high bias conductance, it is also responsible for some of the current flowing through the two topological states at low bias. 
We obtain the maximum value of $e^2/h$ for the zero bias Majorana conductance peak, as expected from an N-S-N setup with symmetric bias drop\cite{Lim-2012,Ulrich-2015,Li-prb2020}. Remarkably, our results show that for a finite chain the value $e^2/h$ for the linear conductance is {\em not} obtained in the whole topological phase, but only near the ``Kitaev points'' ($\mu=0,\,\vert t \vert =\vert \Delta\vert$). Elsewhere the conductance can be close to its maximum value or even significantly lower.

This paper is organized as follows. First, we analyze the spectrum of an isolated Kitaev chain in Sec.~\ref{section: isolated Kitaev chain}, including the higher excitations.
In Sec. \ref{section: transport formulation} we discuss our N-S-N transport setup, providing a general current formula for our system. The analytical expression for the linear conductance is derived in Sec. \ref{section: linear}. In Sec.~\ref{section: non-linear} we present the formula for the differential conductance at finite bias in terms of appropriate Green's functions. The detailed derivations of the expressions for the current, conductance and the Green's functions are given in the Appendices. 
\section{The Isolated Kitaev chain}
\label{section: isolated Kitaev chain}
The central element of our N-S-N system is the finite Kitaev chain, which is a tight-binding chain of $N$ lattice sites, with one spinless fermionic orbital at each site and nearest-neighbor $p$-wave superconducting pairing. The $p$-wave nature of the superconductivity couples particles of equal spin, allowing a spinless treatment. The pairing is treated in the usual mean-field approach, yielding the Kitaev grandcanonical Hamiltonian  \cite{kitaev:physusp2001, Aguado}   
\begin{align}\label{equation: Kit. Hamiltonian, fermionic operators, realspace}
	\hat{H}_{\mathrm{KC}}\,& := \hat H_0 -\mu \hat N_{\mathrm{KC}} =\,-\,t\sum\limits_{j=1}^{N-1}\left(d^\dagger_{j+1} d_{j}+d^\dagger_{j} d_{j+1}			\right)\notag\\
		&\quad +\,\Delta\sum\limits_{j=1}^{N-1}\,\left(d_j^\dagger d_{j+1}^\dagger\,+\,d_{j+1} d_{j}\right)-\mu\sum\limits_{j=1}^N\,d_j^\dagger d_j\,,
\end{align}
in terms of the fermionic creation (annihilation) operators $d_j^\dagger$ ($d_j$). The quantities introduced in Eq. \eqref{equation: Kit. Hamiltonian, fermionic operators, realspace} are the real space position index $j=1,\ldots,\,N$,  the hopping amplitude $t\in\mathbbm{R}$ and the superconducting pairing constant $\Delta\in\mathbbm{R}$. The action of the gate voltage applied later to the wire is to change the chemical potential as $\mu\to \mu +\mu_g$, with $\mu_g=\eta_g e V_g$, and $\eta_g$ the lever arm of the junction.

The spectrum and topological properties of both finite and infinite Kitaev model were discussed in detail in the recent past,~\cite{kitaev:physusp2001, Aguado, Hegde-2015, Zvyagin-2015, Kao-2014, Leumer-2020, Elliott-2015} and we shall give here only a brief overview of the low energy spectrum, giving more emphasis to the hitherto largely unexplored quasiparticle states at higher energy.

In the thermodynamic limit ($N\rightarrow\infty$) the energy of the excitations obeys the bulk dispersion relation 
\begin{align}\label{equation: bulk dispersion relation}
	E_\pm(k)\,=\,\pm\sqrt{\left[\mu\,+\,2t\cos(kd)\right]^2\,+\,4\Delta^2\,\sin^2\left(kd\right)},
\end{align} 
where $d$ is the lattice constant. The topological features of the Kitaev chain can be found after a calculation of the winding number or the Pfaffian topological invariant.~\cite{Review-Chiu-2016, Wen-1989} The boundaries between trivial and non-trivial phases in the topological phase diagram \cite{kitaev:physusp2001, Aguado, Elliott-2015} are determined by the gap closing of the bulk dispersion relation, which happens at $k=0$ or $k=\pi/d$ for $ \Delta\neq0$ and $\mu=\pm 2t$. As one finds, the non trivial phase exists only for $\vert \mu/\Delta\vert < 2\vert t/\Delta\vert$.
\begin{figure}
\begin{center}
	\includegraphics[width=\columnwidth]{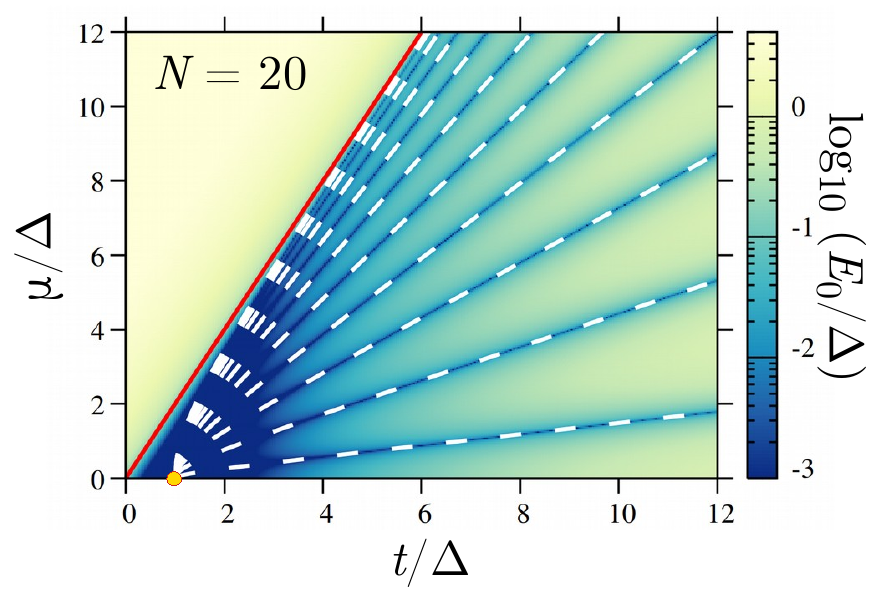}
	\end{center}
	\caption{Ground state energy of the isolated Kitaev chain as a function of $t/\Delta$ and $\mu/\Delta$. 
	The color map in the background displays the numerically calculated energy of the lowest eigenstate, the red line depicts the phase boundary $\vert\mu  \vert=2 \vert t \vert$ between the topologically trivial ($\vert\mu\vert >2 \vert t \vert$) and non-trivial  ($\vert\mu \vert<2 \vert t \vert$) bulk phases, while the white dashed lines are the ``Majorana lines'' defined by Eq. \eqref{equation: MZM condition}. Along these lines the ground state energy $E_0$ is exactly zero. All Majorana lines start  from the $\mu=0$, $t = \pm \Delta$ points and are located in the topological bulk region.}
	\label{figure: groundstate energy and MZM lines}
\end{figure}

In a \emph{finite} chain, the bulk-edge correspondence\cite{Mong-2011,Izumida-2017} implies the existence of evanescent state solutions at the system's boundary in the topologically non-trivial phase. These states have a complex wavevector $\kappa$, and their wave functions decay away from the edges with a decay length $\xi$, which for $\mu=0$ is given by 
\begin{align}\label{equation: decay length}
	\xi = \frac{2 d}{\left\vert \ln \left\vert\frac{t-\Delta}{t+\Delta} \right\vert\right\vert}.
\end{align}
The energy of these topological excitations lies inside the bulk gap introduced with Eq. \eqref{equation: bulk dispersion relation} and is in general non-zero, with the upper bound proportional to $\exp(-Nd/\xi)$; i.e. for $\xi\ll Nd$ the edge state energy is exponentially small.\footnote{ 
The decay length in Eq. \eqref{equation: decay length} is defined for $\mu=0$, since the effect of $\mu$ on $\xi$ is not significant\cite{kitaev:physusp2001, Leumer-2020}.}
The energy of the decaying states becomes \emph{exactly} zero for specific parameter settings\cite{Kao-2014,Zvyagin-2015, Hegde-2015, Kawabata-2017, Leumer-2020}, namely 
\begin{align}\label{equation: MZM condition}
	\mu_n =2\,\sqrt{t^2-\Delta^2}\,\cos\left(\frac{n\pi}{N+1}\right),
\end{align}
with $n=1,\,\ldots,\,N$ and for $t^2\ge\Delta^2$.  Zero energy solutions for  $t^2<\Delta^2$ are found only for $n=(N+1)/2$, which is only possible for odd $N$. 
The zero energy solutions form lines in the $(t/\Delta,\mu/\Delta)$ plane (we shall call them Majorana lines) departing from the points $\vert t \vert=\vert\Delta\vert$, $\mu=0$, as depicted in Fig. \ref{figure: groundstate energy and MZM lines}. 

The exact zero energy solutions of the isolated Kitaev chain represent fermion parity switches,\cite{Beenakker-2013b,DasSarma2012:prb86,Hegde-2015,Pekerten-2019} and for given $t,\Delta$ occur for discrete values of $\mu$. Close to the Majorana lines one always finds eigenstates with exponentially small energy, as seen in Fig. \ref{figure: groundstate energy and MZM lines}. Thus, due to the broadening of the energy levels induced by the coupling to the leads, also states with  energy smaller than such broadening will effectively act as MZM. As we shall show, in an N-S-N setup with symmetric bias they yield a linear conductance very close to $e^2/h$, reaching the exact $e^2/h$ in the thermodynamic limit.\cite{Lim-2012,Ulrich-2015,Li-prb2020} We recall here that in an N-S configuration the height of the zero bias peak is expected to be $2e^2/h$.\cite{Aguado,Prada-2020} 

\subsection{Higher excitation spectrum}

\begin{figure}[htb]
\begin{center}
	\includegraphics[width=0.99\columnwidth]{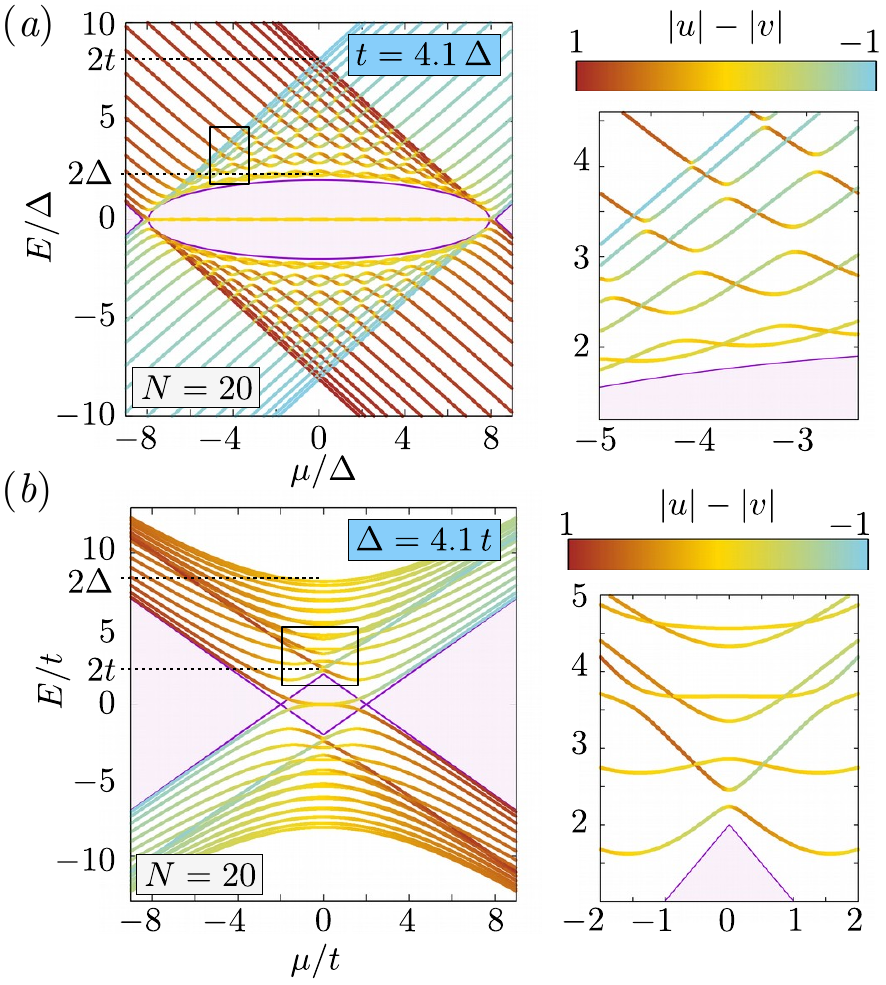}
\end{center}
\caption{\label{figure: full spectrum particle-hole} Full spectrum of a Kitaev chain, for (a) $t=4.1\Delta$ and (b) $\Delta = 4.1t$. The color scale shows the particle/hole character of the corresponding eigenstate, expressed through $|u|-|v|$, where $|u|$ and $|v|$ are the norms of the particle and hole parts of the eigenstate, respectively. The violet shaded regions show the bulk gap.}
\end{figure}

The low energy states of the topological superconductors have garnered so far the most attention of the scientific community. Nevertheless, a current flowing through the Kitaev chain at a larger bias will involve also the higher lying excitations. Thus some questions naturally arise, such as: how will the high energy spectrum impact the differential conductance? If a chain is in the topological phase, will this affect the features visible at finite bias? To answer these questions we first analyze the full spectrum of a finite Kitaev chain. The numerically obtained spectrum as a function of $\mu$ is shown in Fig.~\ref{figure: full spectrum particle-hole}(a) for $t>\Delta$, and in Fig.~\ref{figure: full spectrum particle-hole}(b) for $\Delta > t$. The eigenstates in the Bogoliubov -- de Gennes representation are composed of particle ($u$) and hole ($v$) components. In most of the spectrum the eigenstates have either particle ($|u|>|v|$) or hole ($|v|>|u|$) character; the states within the bulk gap (cf. Appendix~\ref{appendix: bulk gap}), but also some higher energy solutions described below, are nearly equal mixtures of both.

The linear chain, which is the foundation of the Kitaev chain in Eq.~\eqref{equation: Kit. Hamiltonian, fermionic operators, realspace} has inversion symmetry. For the linear chain the inversion corresponds to a straightforward exchange $d^{(\dag)}_{j}\rightarrow d^{(\dag)}_{N+1-j}$, and its matrix representation is an $N\times N$ matrix $I_0$ with 1 on the antidiagonal and 0 elsewhere. If this operation is extended directly to the Kitaev chain, it results in changing the sign of the superconducting pairing because of its p-wave nature. The unitary symmetry inverting the order of the sites, under which the Kitaev Hamiltonian is invariant, is instead $d_{j}\rightarrow i d_{N+1-j}$, $d^{\dag}_{j}\rightarrow -i d^{\dag}_{N+1-j}$. Its matrix representation is $I_{\mathrm{KC}} = i\tau_z \times I_0$, where $\tau_z$ is the Pauli matrix $z$ in the Nambu space. Crucially, $I_{\mathrm{KC}}$ applied to a Nambu spinor adds a global phase $i$ (which can later be gauged away) {\em and} changes the sign of the hole part of the spinor. In consequence, the particle and hole sectors in each eigenstate of the Kitaev chain must have opposite character under the simple inversion symmetry $I_0$ (cf. Fig.~\ref{figure: full spectrum inversion}; for a detailed discussion in a slightly different approach see Appendix~\ref{appendix: excited states}). 

For $t>\Delta$ we see a series of anticrossings between the higher excitations, which occur throughout the spectrum. The particle-like and hole-like solutions of the normal chain in the Nambu space at the anticrossings have the same character under inversion, thus they can hybridize under the influence of the superconducting pairing. In consequence, the particle and hole sectors of the hybridized quasiparticle eigenstates have nearly equal weight. The crossings, on the other hand, are protected by the different inversion symmetries of the involved eigenstates, and have predominantly particle- or hole-like character. For $\Delta>t$ the character of the excitation spectrum is naturally different - higher absolute value of $\mu$ again separates the spectrum into particle- and hole-like sets of states, but at $\mu=0$ the particle-hole mixing occurs within the whole spectrum. Unlike in the $t>\Delta$ case, both the strict and avoided crossings occur now also outside of the topological phase, under the action of the hopping, rather than of the pairing term.\\ 
\begin{figure}[htb]
\begin{center}
	\includegraphics[width=0.99\columnwidth]{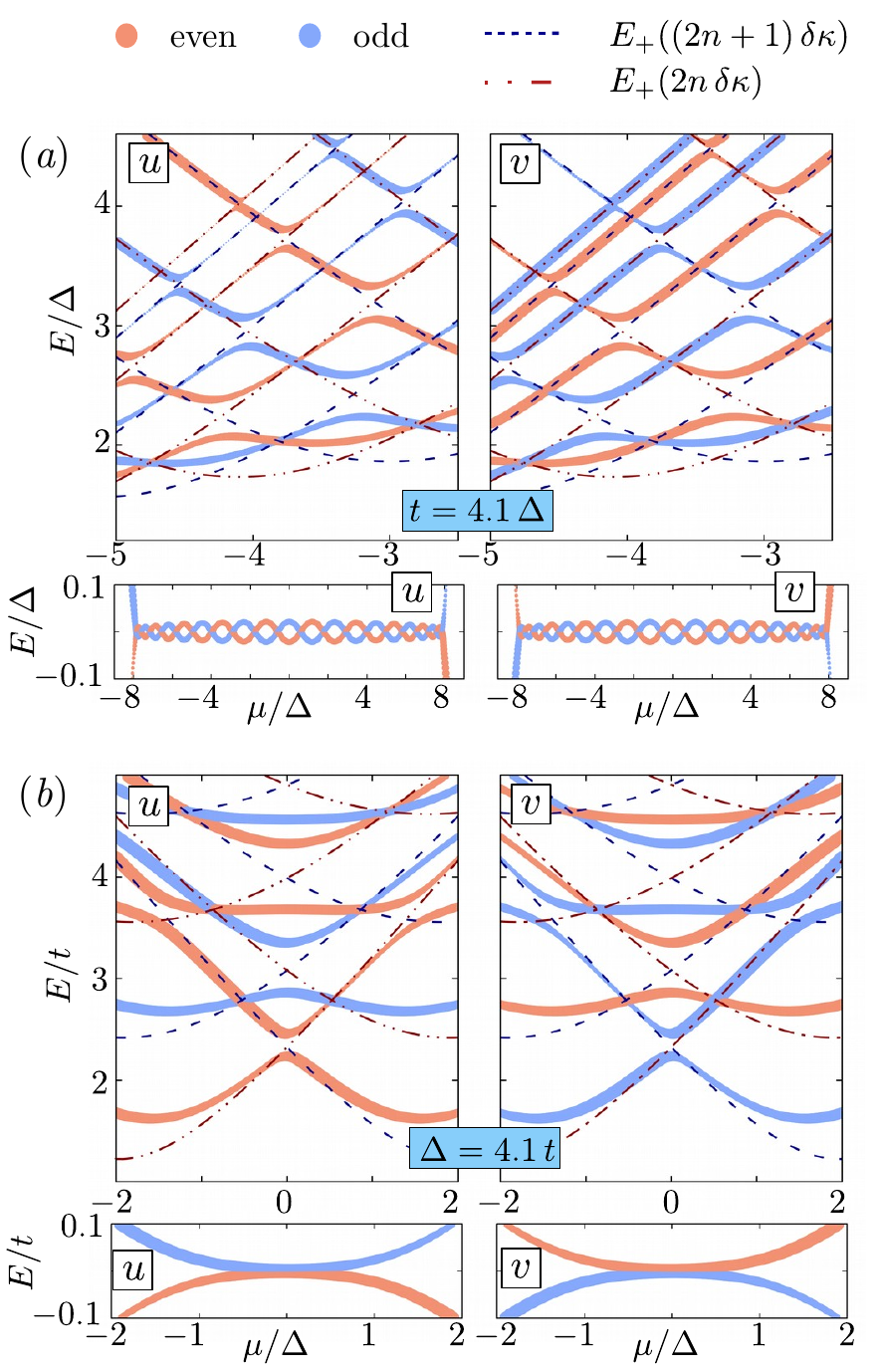}
\end{center}
\caption{\label{figure: full spectrum inversion} Inversion symmetry for chosen ranges of the full spectrum, for (a) $t=4.1\Delta$ and (b) $\Delta = 4.1t$. The color represents $I_u : = \langle u | I_0 | u \rangle /|u|^2$ for the particle sector and $I_v := \langle v | I_0 | v \rangle /|v|^2$ for the hole sector -- light red for $u/v$ even under inversion ($I_{u/v}=+1$), blue for $u/v$ odd under inversion ($I_{u/v} = -1$). Thickness of the lines is proportional to $|u|$ and $|v|$ in the corresponding panel, and the dark dashed and dot-dot-dashed lines follow $E_+((2n+1) \delta\kappa)$ and $E_+(2n\, \delta\kappa)$ from Eq.~\eqref{equation: bulk dispersion relation}, respectively, with $\delta\kappa :=\pi/(N+1)$.
}
\end{figure}

In order to pinpoint the positions of degenerate energies in the spectrum we have to revisit the general quantization rule for the wave vectors of the finite Kitaev chain. As we showed in Ref.~ [\onlinecite{Leumer-2020}], the eigenstates of the Kitaev chain require in general the knowledge of four wave numbers $\pm \kappa_{1,2}$ ($\kappa_1\neq \pm \kappa_2$), since one has to satisfy two boundary conditions for electron and hole sectors separately. We use $\kappa_\Sigma \defl(\kappa_{1}+\kappa_2 )/2$, $\kappa_\Delta\defl(\kappa_{1}-\kappa_2 )/2$ for shortness. The values of $\kappa_{1,2}$ are related and obey
\begin{align}\label{equation: kappa 1 2 depend on mu}
	\cos\left(\kappa_\Sigma \right)\,\cos\left(\kappa_\Delta\right)=-\frac{1}{2}\,\frac{\mu t}{t^2-\Delta^2},
\end{align}
which can be obtained from Eq. \eqref{equation: bulk dispersion relation} by demanding $E(\kappa_1)=E(\kappa_2)$. Thus, Eq. \eqref{equation: kappa 1 2 depend on mu} is in fact a bulk property of the system which also encodes the dependence of $\kappa_\Sigma,\kappa_\Delta$ on the chemical potential in a finite system. Together with the boundary conditions, it yields the quantisation rule of the finite Kitaev chain,
\begin{align}\label{equation: quantisation rule}
	\frac{\sin^2\left[\kappa_\Sigma \left(N+1\right)\right]}{\sin^2\left[\kappa_\Delta \left(N+1\right)\right]}\,=\,\frac{1+\left(\frac{\Delta}{t}\right)^2\,\cot^2\left(\kappa_\Delta\right)}{1+\left(\frac{\Delta}{t}\right)^2\,\cot^2\left(\kappa_\Sigma\right)}.
\end{align}
The description in terms of $\kappa_{\Sigma, \Delta}$ compared with $\kappa_{1,2}$ is more convenient and one can rewrite the dispersion relation into
\begin{align}\label{equation: bulk dispersion relation, seconda versione}
	E^2(\kappa_{\Sigma, \Delta}) &= \frac{1}{\cos^2\left(\kappa_{\Sigma, \Delta}\right)} \left[4(t^2-\Delta^2)\cos^2\left(\kappa_{\Sigma, \Delta}\right)-\mu^2
	\right]\times\notag\\
	&\quad \times\left[\frac{t^2}{t^2-\Delta^2}-\cos^2\left(\kappa_{\Sigma, \Delta}\right)\right],
\end{align}
after the application of Eq. \eqref{equation: kappa 1 2 depend on mu} on Eq. \eqref{equation: bulk dispersion relation}. Moreover, $\kappa_\Sigma,\kappa_\Delta$ define the same state and it can be shown that $E(\kappa_\Sigma) = E(\kappa_\Delta)$ for all values of the parameters  $t$, $\Delta$, $\mu$, $N$. 

At specific values of $\mu$, where the inversion-protected degeneracies occur, $\kappa_\Sigma$ and $\kappa_\Delta$ obey additional constraints. 
We focus first on zero energy crossings, before we turn to the excitations. We have $E(\kappa_{\Sigma})=0$ for $4(t^2-\Delta^2)\cos^2(\kappa_{\Sigma})=\mu^2$. Applying Eq. \eqref{equation: kappa 1 2 depend on mu} to the latter expression  provides $\cos^2(\kappa_{\Delta})=t^2/(t^2-\Delta^2)$ and thus ensures $E(\kappa_{\Delta})=0$. This constraint on $\kappa_\Delta$ is equivalent to $1+(\Delta/t)^2\,\cot^2(\kappa_{\Delta})=0$, which in turn puts a restriction on the quantisation rule, namely $\sin^2\left[\kappa_{\Sigma} \left(N+1\right)\right]=0$, and leads to Eq. \eqref{equation: MZM condition}. The same conclusion can be reached if the first constraint is put on $\kappa_\Delta$.

While a detailed derivation of the position of strict and avoided crossings for $E\neq 0$ can be found in the appendix~\ref{Appendix: energy crossings}, let us here summarize its results. The boundary conditions, together with the requirement of double degeneracy (higher degeneracies occur only for the special cases of either $t=0$ or $\Delta=0$ or $t^2=\Delta^2$), constrain $\kappa_{\Sigma, \Delta}$ to be selected zeros of $\sin^2\left[\kappa_{\Sigma, \Delta} \left(N+1\right)\right]$:
\begin{align}
	\label{equation: crossing, excited states, kappa sigma values}
\kappa_\Sigma &= \frac{n\,\pi}{N+1} \;\textnormal{or}\;\frac{(N+1-n)\,\pi}{N+1}, \quad n=2,\ldots,\,N_{\mathrm{max}},\\
	\label{equation: crossing, excited states, kappa delta values}
\kappa_\Delta &= \frac{m\,\pi}{N+1}, \quad m=1,\ldots,\,n-1,
\end{align}
with $N_\mathrm{max} = N/2$ ($N_\mathrm{max} = (N-1)/2$) for even (odd) $N>3$. These values indeed satisfy the boundary conditions since $\sin^2\left[\kappa_{\Sigma, \Delta} \left(N+1\right)\right]\,[ 1+\left(\Delta /t\right)^2\,\cot^2\left(\kappa_{\Sigma, \Delta}\right)]=0$, for both $t\gtrless\Delta$. For odd $N$ we find additional $(N-1)$ degeneracies at $\mu=0$,~\cite{Leumer-2020} corresponding to $(N-1)/2$ allowed values for $\kappa_\Delta$ if $\kappa_\Sigma=\pi/2$, for both $E\gtrless 0$. The values of $\mu$ where the crossings occur follow from the Eq. \eqref{equation: kappa 1 2 depend on mu}
%
%
for fixed values of $t$ and $\Delta$. While for both $\Delta> t$ and $\Delta <t$ the number of crossings is the same, their positions are not. The energy eigenvalues follow as usual from the dispersion relation in either Eq. \eqref{equation: bulk dispersion relation} or Eq. \eqref{equation: bulk dispersion relation, seconda versione}.\\
The conditions for degenerate energy levels are illustrated in Fig.~\ref{figure: full spectrum inversion}. The energies corresponding to $\kappa_{1,2}=n_{1,2}\,\pi/(N+1)$ are shown with dashed and dot-dot-dashed lines for odd and even $n_{1,2}$, respectively. The conditions \eqref{equation: crossing, excited states, kappa sigma values},\eqref{equation: crossing, excited states, kappa delta values} are obeyed at the intersections of the lines with $n_{1,2}$ either both even or both odd, and indeed at these intersections we see strict crossings. 
On the other hand, the avoided crossing appear for $n_{1,2}$ with different parities. In these cases $\kappa_{\Sigma,\Delta}$ are not integer, but half-integer multiples of $\pi/(N+1)$, and the quantization rule \eqref{equation: quantisation rule} implies
\begin{equation*}
1+\left(\frac{\Delta}{t}\right)^2\cot^2\kappa_\Sigma = 1 + \left(\frac{\Delta}{t}\right)^2\cot^2\kappa_\Delta,
\end{equation*}
which can be fulfilled only if $\Delta=0$ because $\kappa_\Delta \neq \kappa_\Sigma|_{\mathrm{mod}\,\pi}$. Hence, for $\Delta\neq 0$ these crossings are avoided. Interestingly, the values of $E$ and $\mu$ at their centers can be correctly calculated from Eqs.~\eqref{equation: bulk dispersion relation} and \eqref{equation: kappa 1 2 depend on mu} by using $\kappa_{\Sigma,\Delta}$ which are half-integer multiples of $\pi/(N+1)$.\\

One can summarize that the $E\neq 0$ crossings and anticrossings follow the equidistant quantization of a linear chain, where $\Delta=0$ in Eq. \eqref{equation: Kit. Hamiltonian, fermionic operators, realspace}, but the specific values of $\mu$ and the related energies depend on the non-zero $\Delta$. Thus the higher excitation spectrum indeed bears signatures of the topological phase, since at the crossover into the topological phase two of the extended states localize, becoming the boundary modes. The energies and wave functions of the remaining extended states readjust to accomodate the presence of the topological states, although for the extended states this change is continuous.

\section{N-S-N Transport and the current formula}
\label{section: transport formulation}
\begin{figure}[h]
\includegraphics[width=\columnwidth]{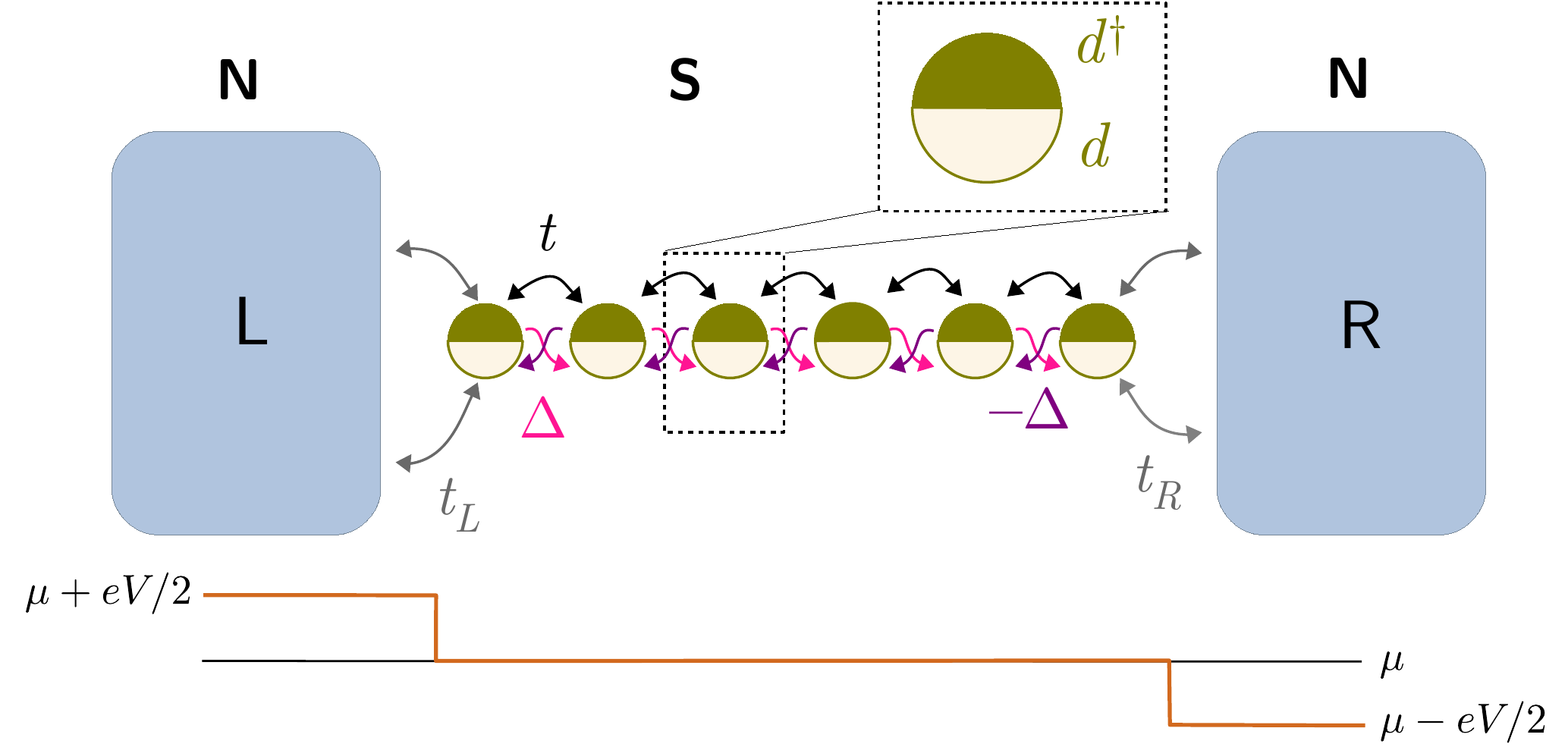} 
\caption{\label{figure: NSN setup}
The Kitaev chain in N-S-N transport configuration, sketched in the particle-hole basis. The parameters $t_L,t_R$ are the tunneling hoppings between the leads and the Kitaev chain from Eq.~\eqref{equation: Tunneling Hamiltonian}. The results shown in this paper are obtained for a symmetrically applied bias. The resulting profile of the chemical potential for the electronic sector is shown below the device.}
\end{figure}

In this section we introduce our transport setup, illustrated schematically in Fig.~\ref{figure: NSN setup}, and discuss the current formula. We place the Kitaev chain between two normal conducting leads, described by  the grandcanonical Hamiltonians 
\begin{align}\label{equation: lead hamiltonian alpha}
	\hat H_\alpha-\mu \hat N_\alpha\,=\,\sum\limits_{k} \epsilon_{k\alpha}\, c_{k\alpha}^\dagger c_{k\alpha},\quad \alpha=L,\,R
\end{align}
where $c_{k\alpha}^\dagger$ $(c_{k\alpha})$ creates (destroys) a spinless fermion in state $k$ and lead $\alpha$. Note that $\hat{H}_{\mathrm{KC}}$ in Eq. \eqref{equation: Kit. Hamiltonian, fermionic operators, realspace} and $\hat H_\alpha$ in Eq. \eqref{equation: lead hamiltonian alpha} are written with the reference energy being the chemical potential $\mu$ of the Kitaev chain. In the above description we consider leads in their eigenbasis. 

Our N-S-N junction is completed with the tunneling Hamiltonian 
\begin{align}\label{equation: Tunneling Hamiltonian}
\hat 	H_\mathrm{T}\,&=\,\sum\limits_{k}\,\left(t_L\, c_{kL}^\dagger d_1\,+\,t_L \,d_1^\dagger\,c_{kL}\right),\notag\\
		&\quad +\,\sum\limits_{k}\,\left(t_R \,c_{kR}^\dagger \,d_N \,+\,t_R\,d_N^\dagger\,c_{kR}\right),
\end{align}
which couples only the first (last) chain site to the contact $L$ $(R)$. We consider the tunneling elements $t_{L,R}$ as $k$-dependent, real quantities.
This setup is equivalent to a fully spin polarized system where the fixed spin $\sigma$ is suppressed in the notation. 
The current through an N-S-N junction can be calculated with the NEGF approach
\cite{Jauho,Flensberg,Ryndyk,Rammer-1986,Meir-Wingreen-1992,LevyYeyati-1995,diVentra,Zeng-2003,Sriram-2019}. Since the Kitaev Hamiltonian in Eq. \eqref{equation: Kit. Hamiltonian, fermionic operators, realspace} is given in mean field, and thus breaks the conservation of particles, the calculation has to be carried out with care. The physical current measured in experiment is conserved everywhere, thus we calculate it for simplicity inside the left lead. There, the electronic current (for fixed spin) reads,
\begin{align}
	I_{L}(t)\,=\,-e \,\langle \dot{N_{L}}\rangle,
\end{align}
where $e$ is the elementary charge and $N_{L} = \sum\limits_k c_{kL}^\dagger c_{kL}$.  The steady state current (for fixed spin) reads
\begin{align}\label{equation: current formula bdg basis}
	I_L&=i\frac{e}{2h}\int\limits_{\mathbbm{R}}dE\,\mathrm{Tr}\left\{\tau_z \otimes \mathbbm{1}_N \, \Gamma_L
		\left[ G^<+F_L\,\left( G^r - G^a\right)		\right]\right\},
\end{align}
with $\Gamma_\alpha = -2 \,\mathrm{Im}\left(\Sigma_\alpha^r\right)$ ($\alpha = L,R)$, where $\Sigma_\alpha^r$ is the retarded self-energy of the lead $\alpha$, defined in Appendix~\ref{Appendix: Derivation of the current formula}. Since we are working in the BdG formalism, all quantities under the trace in the above equation are $2N\times 2N$ matrices, defined w.r.t $\hat{\Psi}=\left(d_1,\,\ldots,d_N,\,d_1^\dagger,\ldots,\,d_N^\dagger\right)^\mathrm{T}$, forming electronic and hole subspaces. For example, the matrix $F_\alpha$ contains the Fermi functions $f(E)$ for electrons and holes in the following form
\begin{align}
	F_\alpha\,=\,\left[\begin{matrix}
	\mathbbm{1}_N\,f(E-eV_\alpha) &\\
	& \mathbbm{1}_N\,f(E+eV_\alpha) 
	\end{matrix}
	\right],
\end{align}
where $V_L = \eta V$, $V_R = (\eta-1)V$ account for scenarios with different applied bias $V$. Further, the lesser Green's function $G^<$ is 
\begin{align}
	G^<\,=\,i\,G^r \,\left(\sum\limits_{\alpha=L,R}F_\alpha\Gamma_\alpha\right)\,G^a,
\end{align}
with details of the derivation discussed in appendix \ref{Appendix: Derivation of the current formula}. In equilibrium $(V=0)$ we have $G_\mathrm{eq}^< = -F(E)\,(G^r-G^a)$ and the current vanishes.

The special choice of the tunneling Hamiltonian in Eq. \eqref{equation: Tunneling Hamiltonian} defines the self-energies $\Sigma_\alpha^{r}$ as sparse matrices, see Eqs. \eqref{appendix equation: self energy Sigma L}, \eqref{appendix equation: self energy Sigma R}. This, together with the trace and the particle-hole symmetry, yields a current formula where only two entries of the retarded Green's function, namely $G^r_{1,N}$ and $G^r_{1,N+1}$, are required. One finds
\begin{align}\label{equation: current formula, two processes}
	I_L&= \frac{e}{h}\int\limits_{\mathbbm{R}} dE \left\{\Gamma_L^- \Gamma_R^-\,\left\vert G^r_{1,N}\right\vert^2\,\left[f(E-eV_L)-f(E-eV_R)\right]			\right.\notag\\
	 &\qquad + \left. \Gamma_L^-\,\Gamma_L^+\, \left\vert G^r_{1,N+1}\right\vert^2\,\left[f(E-eV_L)-f(E+eV_L)\right]\right\},
\end{align}
now setting $V_L=-V_R=V/2$. We choose this scenario to keep the current in Eq. \eqref{equation: current formula, two processes} conserved, $I_L=-I_R$, which for symmetric bias occurs if $\Gamma_L=\Gamma_R$, even without a self-consistent calculation of $\Delta$.\cite{LevyYeyati-1995,Melin-2009,Lim-2012}
The density of states in the lead $\alpha$ and the associated tunneling amplitudes $|t_\alpha(k)|^2$ are encoded in the quantities
$\Gamma_\alpha^\pm =  2\pi \sum\limits_k\,\vert t_\alpha(k)\vert^2\,\delta(E\pm\epsilon_{k\alpha})$, with $-$ ($+$) for particles (holes). In a realistic device scenario one may have however to represent the leads in the site basis and employ a recursive approach to calculate the self energy.~\cite{Sriram-2019}

Eq. \eqref{equation: current formula, two processes} allows a microscopic analysis of the charge transfer through the Kitaev chain, where two processes contribute. The term containing $G^r_{1,N}$ describes the usual direct transfer ($D$) of a quasiparticle from the left to the right lead through a normal conducting system, but here in presence of the p-wave superconductivity embodied by $\Delta$. The second term in Eq. \eqref{equation: current formula, two processes}, i.e. the one including $G^r_{1,N+1}$, describes the Andreev reflection -- the incoming electron is reflected back as a hole and a right moving Cooper pair is formed inside the Kitaev chain.\cite{Andreev-1964,Blonder-1982}. In the third possible process the right-moving Cooper pair in the chain is formed by an electron coming from the left and a hole coming from the right. This process, named crossed Andreev reflection, does not contribute to the current in a symmetric bias configuration.
We give the exact analytic form of $G^r_{1,N}$ and $G^r_{1,N+1}$ in appendix \ref{appendix: exact expression GF for current} in terms of Tetranacci polynomials, see in particular Eqs.~\eqref{appendix equation: retarded GF Andreev},\eqref{appendix equation: retarded GF direct}.

The relative weight of the two contributing processes depends on the chosen parameters of the Kitaev chain ($\mu,\,t,\Delta$), as we will see in the context of the zero temperature conductance in the next section.

%
%
%
%
\section{Linear transport}
\label{section: linear}
The conductance $G\defl \lim\limits_{V\rightarrow0} \partial I/\partial V$ is easily calculated from  Eq. (\ref{equation: current formula, two processes}). At $T=0 K$, one finds the simple formula
\begin{align}\label{equation: conductance formula, three processes}
	G &= \frac{e^2}{h}\left\{\Gamma_L^- \Gamma_R^-\,\left\vert G^r_{1,N}\right\vert^2  +		
 \,\Gamma_L^-\,\Gamma_L^+\, \left\vert G^r_{1,N+1}\right\vert^2 \right\}_{E=0}\notag \\
	&=: G_D+G_A,
\end{align}
accounting for direct transport and Andreev reflection, respectively.  

In the following we make use of the  analytic expressions for $G^r_{1,N}$, $G^r_{1,N+1}$ derived in Appendix \ref{appendix: exact expression GF for current} to give the closed formulas for $G^r_{1,N}$, $G^r_{1,N+1}$ at $E=0$ (derived in Appendix \ref{appendix: Conductance formula}). For simplicity we consider the wide band limit, where the tunneling amplitudes $t_L,t_R$ and the densities of states $\rho_L,\rho_R$ in the leads are constant. Thus, $\Gamma_{L,R}^\pm =\Gamma_{L,R} =const$. We find that the Green's functions at $E=0$ are given by
\begin{align}
\left. G^r_{1,N}\right|_{E=0} &= (-1)^{N-1}\, \frac{p^{N-1} + m^{N-1}}{|q_+|^2 + \gamma_L\gamma_R(p^{N-1}+m^{N-1})^2} \,q_-, \\[2mm]
\left. G^r_{1,N+1}\right|_{E=0} &= -i\gamma_R\, \frac{p^{2N-2} - m^{2N-2}}{|q_+|^2 + \gamma_L\gamma_R(p^{N-1}+m^{N-1})^2},
\end{align}
with $p= t+\Delta$,  $m = t-\Delta$ and $\gamma_{L,R} = \Gamma_{L,R}/2$. The polynomial $q_s$ ($s = \pm 1$), given by 
\begin{align}\label{equation: definition of q_s}
	q_s\,=\,p^{N-2}\,&\left[s\,p^2\,x_{N,0}\,+\,ip\,x_{N-1,0} (s\,\gamma_L-\gamma_R)\right.\notag\\
	&\quad + \left.\,x_{N-2,0}\, \gamma_L\,\gamma_R	\right]\,,
\end{align}
carries information on the spectral structure of the isolated Kitaev chain, since the determinant of the   isolated  Kitaev Hamiltonian with $N$ sites is $(-1)^N\,p^{2N}\,x_{N,0}^2$.~\cite{Leumer-2020}  
The closed form of the term $x_{j,0}$ for an arbitrary integer $j$ is 
\begin{align}
	x_{j,0}\,=\,\frac{R_+^{j+1}-R_-^{j+1}}{R_+ -R_-}\,, 
\end{align}
with $R_\pm=(-\mu\pm\sqrt{\mu^2-4 \,mp})/(2p)$. \\

We find for the conductance the closed form 
\begin{align}\label{equation: Conductance, wide band limit zero temperature}
G=\,\frac{e^2}{h}~\,\frac{\gamma_L\,\gamma_R \left(p^{N-1}+m^{N-1}\right)^2}{\vert q_+\vert^2\,+\,\gamma_L\,\gamma_R \left(p^{N-1}+m^{N-1}\right)^2} \,.
\end{align}
\begin{figure}
	\includegraphics[width=\columnwidth]{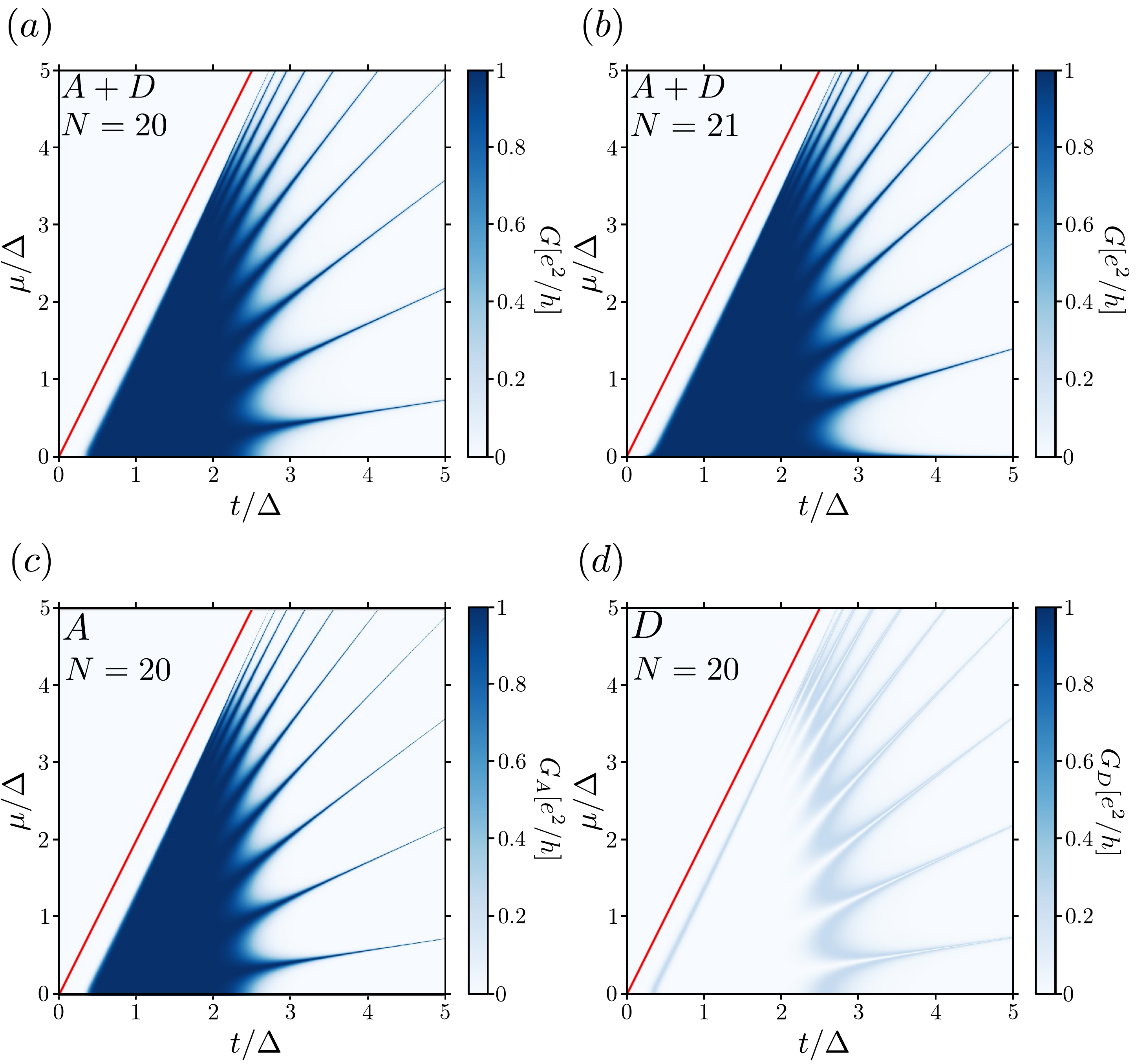}
	\caption{Conductance $G$ (wide band limit) in units of $e^2/h$ for $\gamma_{L,R}/\Delta = 0.001$ as function of $\mu/\Delta$ and $t/\Delta$. (a),(b) The roughly triangular plateau of high conductance is bounded (with some padding\cite{Leumer-2020}) by the phase boundary (red line) and branches out into distinct lines when the magnitude of decay length and system length become comparable. Those lines of high conductance follow the Majorana lines given by Eq. \eqref{equation: MZM condition}, with one of them always coinciding with the $\mu=0$ axis for odd $N$. (c) The most important contribution to the conductance $G=G_A+G_D$ is the Andreev term. (d) The direct term $G_D$ only broadens the conductance plateau.}
	\label{figure: Conductance for even/ odd N and andreev/ direct contributions}
\end{figure}
The conductance in the limit $N\rightarrow\infty$ takes the value $e^2/h$. We also get the value $G= e^2/h$ for the linear conductance at the Kitaev points, independent of the value of the coupling strengths $\gamma_{L,R}$,   since the terms in $q_s$ vanish there. The $G_L$ from equations (32),(33) in Ref.~\onlinecite{Stoof-Kitaev} can be obtained from our Eq.~\ref{equation: Conductance, wide band limit zero temperature} by setting $\mu=0$ and adding the factor 2 (in Ref.~\onlinecite{Stoof-Kitaev} one contact is effectively grounded). 

Besides the Kitaev points the behavior of the conductance is more intricate and depends  on the parameters setting. In particular, on the zero energy Majorana lines of the isolated chain, see Eq. \eqref{equation: MZM condition},
the term $x_{N,0}$ vanishes, although, due to the coupling to the leads, the whole polynomial $q_s$ does not.  For the special case of symmetric coupling $\gamma_L=\gamma_R$ the conductance along the Majorana lines becomes however nearly independent of the coupling.   
The behavior of the conductance in the $t/\Delta$ - $\mu/\Delta$ plane is shown  in Fig. \ref{figure: Conductance for even/ odd N and andreev/ direct contributions} a),b) for the case of $N=20$ and $N=21$ sites. 
While in the vicinity of the Kitaev points the conductance is large and close to $e^2/h$, as the ratio of $t/\Delta$ increases it remains so large only in close vicinity of the Majorana lines. 

In order to better understand this behavior, we examine more closely the two contributions to the conductance, $G_D$ and $G_{A}$, see Eqs~\eqref{equation: conductance formula, three processes}. We find
\begin{align}\label{equation: direct conductance term}
	G_D\,&=\,\frac{e^2}{h}\,\frac{\gamma_{L}\,\gamma_R\left(p^{N-1}+m^{N-1}\right)^2}{\left[\vert q_+\vert^2 +\gamma_{L}\,\gamma_{R}\left(p^{N-1}+m^{N-1}\right)^2\right]^2}~
	\vert q_- \vert^2,\\
	\label{equation: Andreev conductance term}
	G_A\,&=\,\frac{e^2}{h}\,\frac{\gamma^2_{L}\,\gamma^2_R\left(p^{2N-2}-m^{2N-2}\right)^2}{\left[\vert q_+\vert^2 +\gamma_{L}\,\gamma_{R}\left(p^{N-1}+m^{N-1}\right)^2\right]^2},
\end{align}
with $q_\pm$ from Eq. \eqref{equation: definition of q_s}. For details of the calculation, see appendix \ref{appendix: Conductance formula}. The contributions $G_A$ and $G_D$ for the case $N=20$  are depicted in Fig. \ref{figure: Conductance for even/ odd N and andreev/ direct contributions} c), d). The difference between the Andreev and the direct term originates from the function $q_-$ which appears in the numerator of $G_D$. For $\gamma_{L,R} \ll \Delta$ the $q_-$ factor is small as long as $\xi \ll dN$, i.e. inside the triangular conductance plateau. Here $x_{N,0}$ is exponentially small due to the existence of in-gap states and $x_{N-1,0}$, $x_{N-2,0}$ are suppressed by $\gamma_{L,R}$. In the region of the plateau $q_+$ is also small, thus $G_A$ is enhanced while $G_D$ is suppressed. In the limit of vanishing order parameter, $\Delta=0$, it immediately follows from the above equations that $G=G_D$, $G_A=0$, since -- as expected -- the Andreev contribution vanishes. For $\xi \gg dN$ and leaving the discrete lines of non-zero conductance aside for a second, we find no in-gap states with zero, or even exponentially small energy anymore; the function $q_s$ grows for increasing values of $\mu/\Delta$ and/ or $t/\Delta$, which leads to a suppression of both conduction terms $G_{D} \propto \vert q_-\vert^2/\vert q_+\vert^4$, $G_{A} \propto 1/\vert q_+\vert^4$. For intermediate parameter values $\xi \approx dN$, the polynomials $x_{N-1,0}$, $x_{N-2,0}$ become important. They describe essentially the spectrum of a Kitaev chain with $N-1$, $N-2$ sites, i.e. $\xi \gtrapprox d(N-j)$ for $j=1,2$. Their contributions define the crossover region between the triangular plateau of high conductance and the region featuring separated Majorana lines within the topologically non-trivial phase when $\xi \gg dN$. Note that the crossover region is influenced by $\gamma_{L,R}$ too.

\begin{figure}[hbt]
	\centering
	\includegraphics[width = \columnwidth]{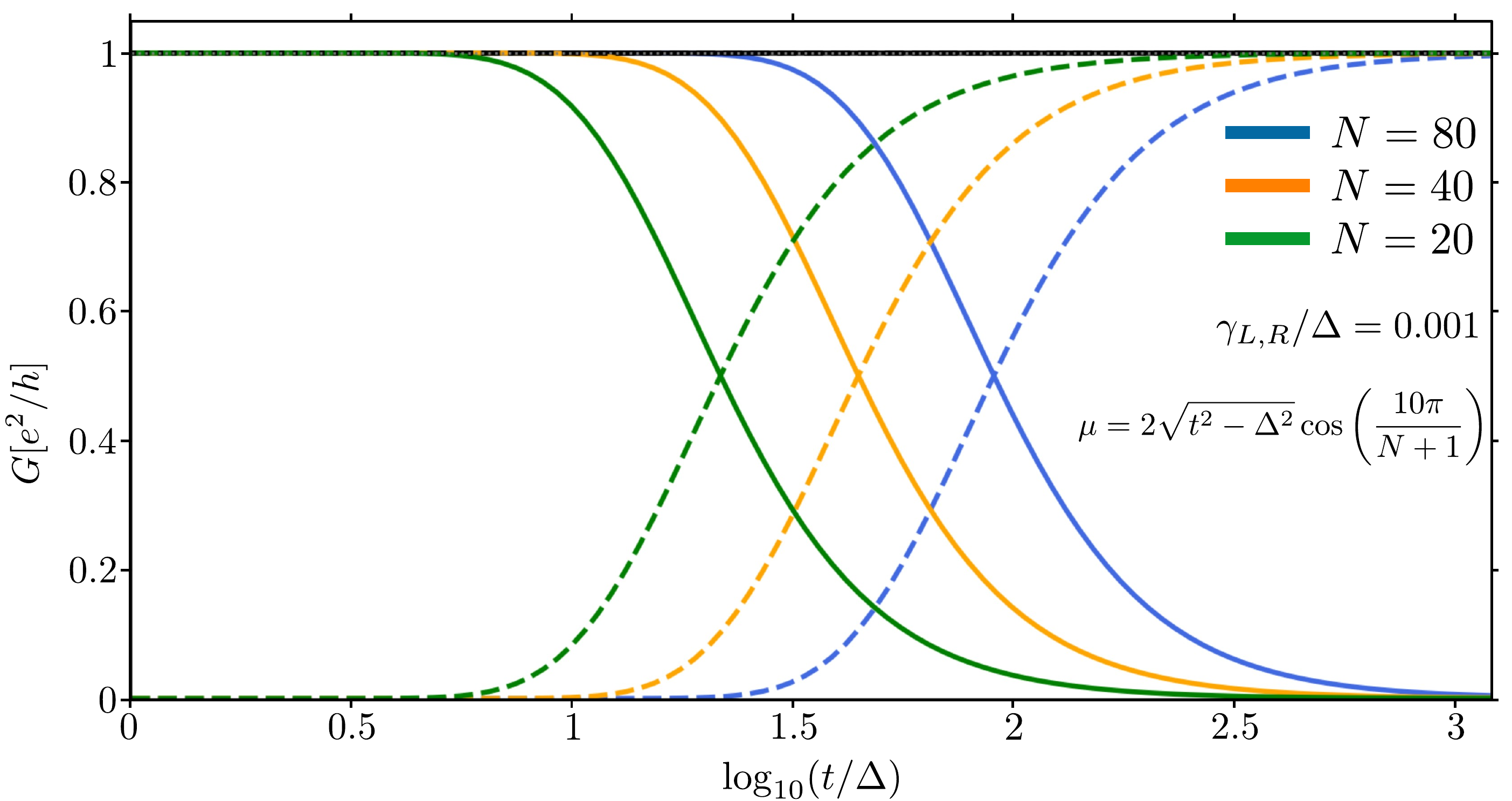}
	\caption{Conductance contributions $G_A$ and $G_D$ (wide band limit, $T=0\,$K) in units of $e^2/h$ as function of $t/\Delta$ along the zero energy line, with $\mu$ adjusted to obey Eq. \eqref{equation: MZM condition} for $n=10$ and with $\gamma_L=\gamma_R = 0.001\Delta$, for different chain lengths. The Andreev term (solid lines) mostly contributes in the vicinity of the Kitaev point and decreases for larger ratios $t/\Delta$, while $G_D$ (dashed lines) shows the opposite behavior. The total conductance (black line) $G=G_A +G_D$ stays close to $e^2/h$, since a contributing zero energy eigenstate of the isolated Kitaev chain is always available. The Andreev term accounts here for the reflection $R_D=1-T_D$, where $T_D$ is the transmission amplitude of the direct term.}	
\label{figure: conductance on MZM lines}
\end{figure}

Let us turn to the conductance along the Majorana lines, given by \eqref{equation: MZM condition}. On those lines the function $x_{N,0}$ vanishes and thus the functions $q_s$ have minima in $\mu$. The value of $q_s$ varies strongly around these minima and leads to the appearance of low conductance regions between the Majorana lines for $\xi \gg dN$. The ratio of $G_D$ and $G_A$ changes along those lines as depicted in Fig. \ref{figure: conductance on MZM lines} starting with $G_A = 1$ and $G_D=0$ at Kitaev points and converges to $G_A=0$ for $t/\Delta \rightarrow \infty$. This behavior is independent of the chosen line. Remarkably, the sum $G_D+G_A$ is seemingly constant and equal to $e^2/h$; it is in fact very slightly suppressed due to Eq. \eqref{equation: Conductance, wide band limit zero temperature}, becoming fully quantized only in the thermodynamic limit. 
\section{Non-linear transport}
\label{section: non-linear}
\begin{figure}[hbt]
	\centering
	\includegraphics[width = \columnwidth]{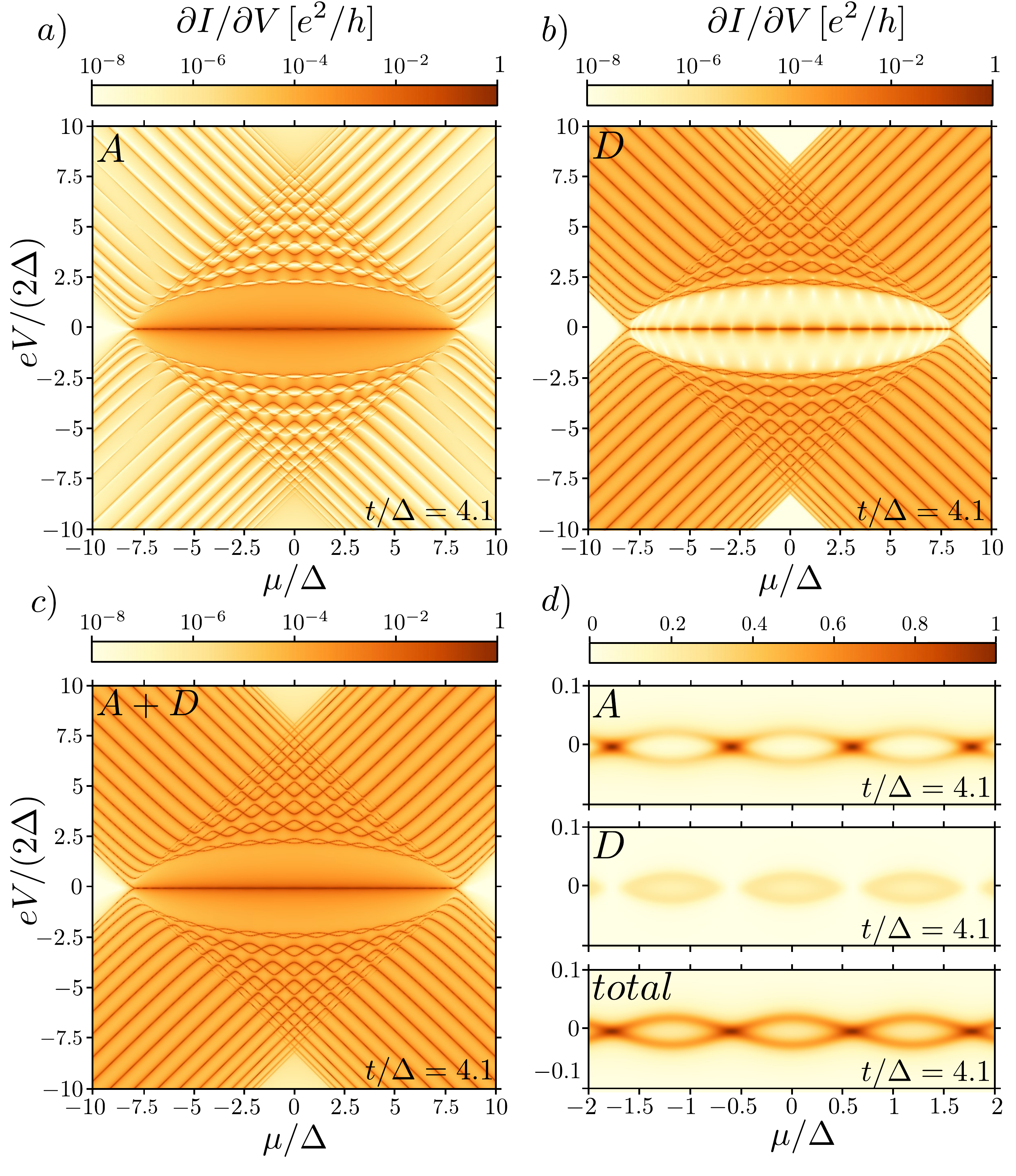}
	\caption{Differential conductance as a function of $eV/(2\Delta)$ and $\mu/\Delta$ for $\vert t /\Delta \vert = 4.1$, $\gamma_L=\gamma_R=0.02\Delta$, and $N=20$. (a) The Andreev term is the dominant contribution to the differential conductance for the in-gap states, but also affects the excitations. (b) The direct contribution to $\partial I /\partial V$ is present for all eigenstates, its strength inside the gap depends on the parameters. (c) The total differential conductance is the sum of the Andreev and the direct terms. (d) The dark stripe for $V\approx 0$ of the Andreev term in (a) has in fact a braid-like structure, since the chain is too short ($Nd\sim 4\xi$) to support zero energy eigenstates everywhere.\cite{Aguado, Leumer-2020, kitaev:physusp2001} The Andreev reflections are stronger around the values of $\mu$ where exact zero energy states are present. Direct process contributions enhance the transport between two Andreev peaks.}
	\label{figure: differential conductance t/d = 4.1}
\end{figure}

\begin{figure}[hbt]
	\centering
	\includegraphics[width = \columnwidth]{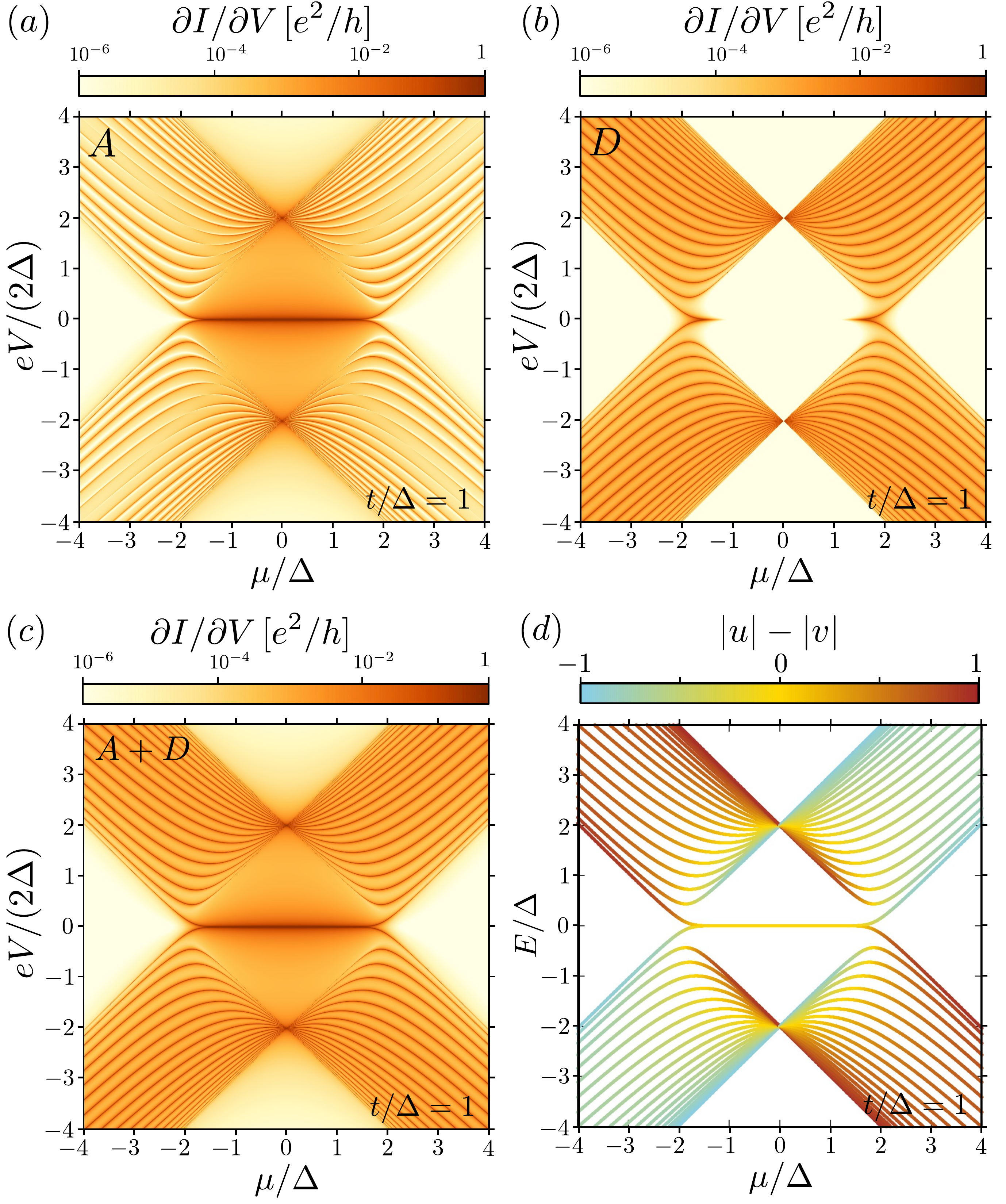}
	\caption{Differential conductance as a function of $eV/(2\Delta)$ and $\mu/\Delta$, with $\gamma_L=\gamma_R=0.02\Delta$ and $N=20$, similar to Fig. \ref{figure: differential conductance t/d = 4.1}, but for $\vert t\vert=\vert\Delta \vert$. (a) The Andreev term is still dominant around $V\approx 0$, while it is weak for excited states except at $\mu=0$ and $V=\pm 2\Delta$. (b) The direct contribution shows the complementary behavior to the Andreev one. Further, it does not contribute at $(\mu,V)=(0,\pm 2\Delta)$. (c) Total differential conductance as sum of the Andreev and direct term shows a higher conductance inside the gap and at $\mu=0$ and $V=\pm 2\Delta$. (d) The spectrum of the isolated Kitaev chain, where the brown (blue) color indicates the particle (hole) character as in Fig.~\ref{figure: full spectrum particle-hole}. The Andreev term in a) contributes more strongly for states with equal particle and hole parts.}
		\label{figure: differential conductance t/d = 1}
\end{figure}

The non-linear transport effects are captured by the differential conductance $\partial I/ \partial V$. At $T=0\,$K and using Eq. \eqref{equation: current formula, two processes} we find
\begin{align}\label{equation: differential conductance zero temperature}
	\frac{\partial I}{\partial V}&= \frac{e^2}{2h}\,\sum\limits_{E=\pm V/2}\Gamma_L^-\left(
	 \Gamma_R^-\,\left\vert G^r_{1,N}\right\vert^2+\Gamma_L^+\, \left\vert G^r_{1,N+1}\right\vert^2
	\right),
\end{align}
where we set $V_L=-V_R = V/2$. We depicted $\partial I/ \partial V$ and its Andreev (A) and direct (D) contributions given by the $G^r_{1,N+1}$ ($G^r_{1,N}$) terms in Figs. \ref{figure: differential conductance t/d = 4.1} and \ref{figure: differential conductance t/d = 1}.

As expected, the Andreev term is slightly smaller than $e^2/h$ around $V\approx 0$ for $\vert \mu \vert <2\vert t\vert$ and $\xi/(dN)\ll 1$, while the direct term is weak. Outside $V\approx 0$ the roles of Andreev and direct contributions are exchanged, though the Andreev term reemerges at the resonances with the quasiparticle energy levels and inside the avoided crossings between higher excitations; there the involved eigenstates of the Kitaev chain have again significant contributions from both particle and hole sectors (cf. Fig. \ref{figure: full spectrum particle-hole}). 

A special situation arises at the Kitaev points, where $\mu=0$ and $\vert t \vert =\vert \Delta \vert$. Here the isolated Kitaev chain hosts only eigenstates with energies $0,\,\pm 2 \vert t\vert$ (degenerate). For these parameters direct charge transfer through the Kitaev chain is forbidden. This becomes evident when the Kitaev chain Hamiltonian is represented in terms of Majorana operators (see Fig.~\ref{figure: isolated - Majorana}), where one of the nearest neighbor hopping amplitudes, either $i(\Delta+t)$ or $i(\Delta-t)$ vanishes, and the chain falls apart into a set of dimers and two end sites.\cite{Aguado,kitaev:physusp2001, Leumer-2020} The direct term $G^r_{1,N}$ cannot contribute to transport, which occurs only through the Andreev term $G^r_{1,N+1}$ | as long as $\Gamma_R\neq 0$, the Cooper pair formed in the Kitaev chain through the Andreev reflection can escape into the right lead. When $\mu\neq 0$, the chemical potential binds Majorana operators of the same site and establishes a direct transport channel linking the dimers and end sites. 

Let us turn back to the region around $V\approx 0$ for  $\vert \mu \vert <2\vert t\vert$, $\xi/(dN)\ll 1$ and $\vert \Delta\vert \neq \vert t \vert$. The seemingly structureless Andreev contribution to $\partial_V I$ in Fig.~\ref{figure: differential conductance t/d = 4.1}(a) has in fact a braid-like pattern of larger values as depicted in panel (d) of Fig. \ref{figure: differential conductance t/d = 4.1}. The higher values $\approx e^2/h$ for the Andreev term arise around the $\mu$ values where the MZM are present~\cite{Kao-2014,Hegde-2015,Zvyagin-2015,Leumer-2020}, i.e. at the zero-energy crossings. In between these specific parameter values the importance of the Andreev contribution decreases and the direct term starts to contribute.


\section{Conclusions}
In this work we have investigated linear and non-linear transport across a finite Kitaev chain in an N-S-N setup, with symmetrically applied bias.
Using the analytical methods developed to study the spectrum of the isolated Kitaev chain\cite{Leumer-2020}, we could provide closed formulae for the 
relevant Green's functions and in turn for the linear and differential conductance at zero temperature. We have analyzed the quasiparticle spectrum with its complex pattern of strict and avoided crossings being governed by the inversion symmetry, and related this pattern to the calculated transport spectra. Perhaps counterintuitively, our results show that also direct transmission processes contribute to the subgap transport mediated by the topological states, and likewise, that the Andreev processes participate in the transport at high bias, especially when the involved states are nearly equal superpositions of particle and hole solutions. Further, remarkably, in a finite chain even along the Majorana lines the linear conductance is only approaching its maximum value of $e^2/h$, reaching it only near the Kitaev points. 

In summary, our work provides a complete description of transport through an archetypal topological superconductor, extending our knowledge of this system beyond what can be gleaned from minimal models reduced to topological states alone. Since some of the observed spectral and transport features are generic to 1D topological superconductors, our complete analytical and numerical treatment can provide a valuable benchmark and insight for the study of other model systems, such as for example the one based on s-wave proximitized Rashba nanowires.

\acknowledgments{
NL and BM thank for financial support the Elite Netzwerk Bayern via the IGK "Topological Insulators" and the Deutsche Forschungsgemeinschaft via SFB 1277 Project B04. 
BM would like to acknowledge funding from the Science and Engineering Research Board (SERB), Government of India under Grant No. STR/2019/000030, and the Ministry of Human Resource Development (MHRD), Grant no. STARS/APR2019/NS/226/FS under the STARS scheme.
}

\FloatBarrier
\appendix

\section{Bulk gap}
\label{appendix: bulk gap}
\begin{figure}[htbp]
\includegraphics[width=\columnwidth]{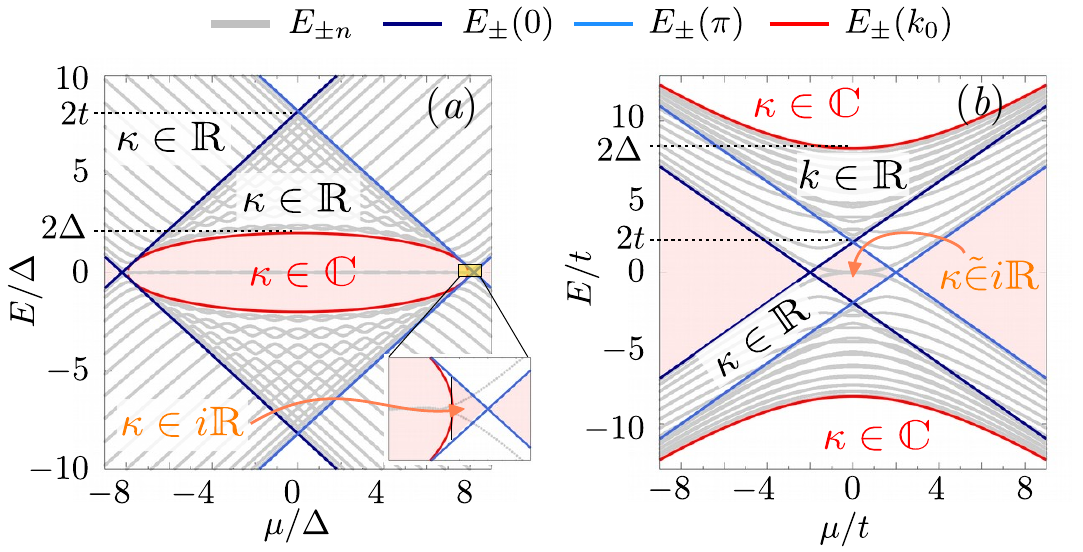} 
\caption{
\label{figure: bulk gap}
The band extrema of a bulk Kitaev chain as a function of $\mu$, for (a) $t=4.1\Delta$ and (b) $\Delta=4.1t$. The numerical energy levels for a chain with $N=20$ sites are shown in grey for comparison.
The types of allowed solutions for the wave number $\kappa$ in a finite chain are indicated. Purely imaginary $\kappa$ are allowed in the small range of $\mu$ indicated in the inset of (a). The notation $\kappa\tilde{\in} i\mathbb{R}$ in the topological range of $\mu$ in (b) means that the allowed wavevectors have also a constant real part, $\pi/2$.
}
\end{figure}

The bulk gap in the Kitaev spectrum is easily estimated from the dispersion relation \eqref{equation: bulk dispersion relation}. The condition for the vanishing first derivative at the band extrema is fulfilled at three values of bulk momentum $k$,  
\begin{equation*}
 k_1 = 0,\quad k_2 = \pi,\quad k_0 = \arccos \left(\frac{\mu t}{2(\Delta^2-t^2)}\right).
\end{equation*}
Examples of the spectra of the Kitaev chain with $N=20$ sites as a function of $\mu$, together with the lines denoting the bulk energies at the band extrema, are shown in Fig.~\ref{figure: bulk gap}. The wave functions in a finite Kitaev chain can be described by purely real, purely imaginary or complex wave vectors $\kappa$, in the regions marked in the figure. The bulk gap is marked with light red shading.

\section{Selected quantum bases for the Kitaev chain}

\begin{figure}[h]
\begin{center}
\includegraphics[width=0.9\columnwidth]{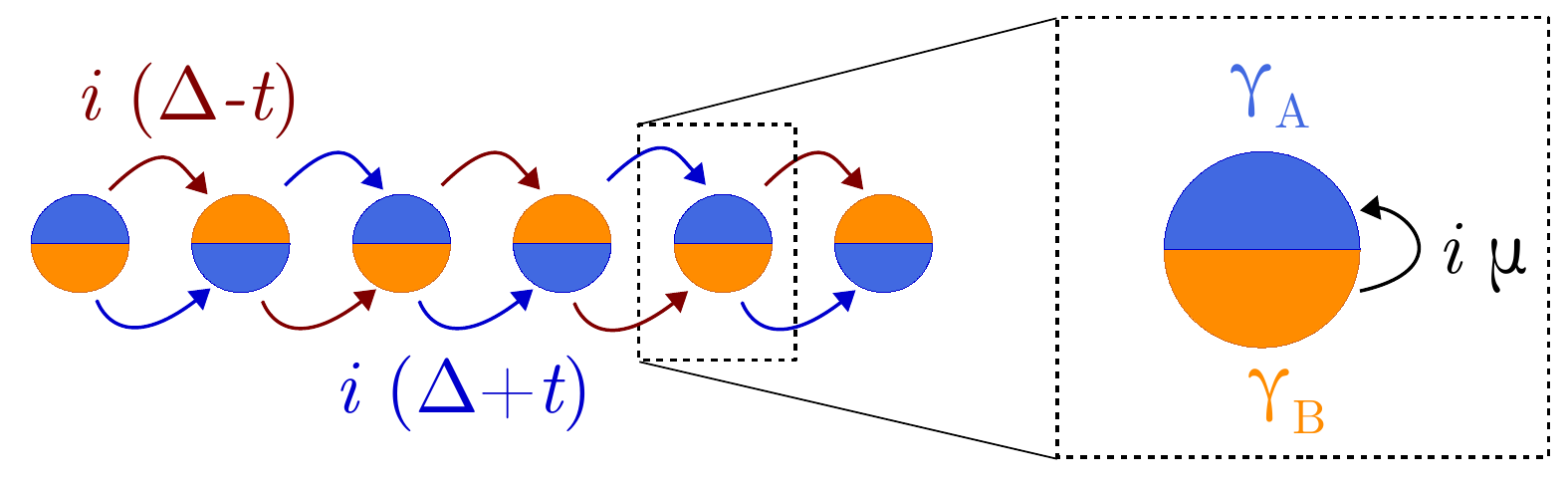} 
\end{center}
\caption{\label{figure: isolated - Majorana}
Kitaev chain described through the Majorana operators. The effective hoppings are all imaginary, with $i(\Delta\pm t)$ describing the hopping between two sites and the $i\mu$ relating the Majorana $A$ and $B$ sublattices on one site.}
\end{figure}

The operators associated with the Kitaev chain can be represented in several bases, each suited to facilitate some specific calculation. We give here an overview of the four basis choices which will be used in these Appendices, together with the rationale behind this choice.
\begin{enumerate}
\item {\em Default Bogoliubov - de Gennes basis}, used throughout the main text and given by
\begin{equation}
\label{appendix equation: default basis}
\hat{\Psi} = (d_1,...,d_N,d_1^\dag,...,d_N^\dag)^T. 
\end{equation}
This basis neatly separates particle and hole sectors of the system.
The representations of the physical quantities in this basis do not carry any labels: the Green's functions are $G^{s}$, where $s=<,>,a,r$, the self-energies $\Sigma^{s}_\alpha$, the $\Gamma$ matrices $\Gamma_\alpha$ and $\alpha = L,R$.
 \item {\em Chiral basis} is defined by
\begin{equation}
\label{appendix equation: chiral basis}
\hat{\Psi}_c = (\gamma_1^A,...,\gamma_N^A,\gamma_1^B,...,\gamma_N^B)^T, 
\end{equation}
where $\gamma^{A/B}$ are the Majorana operators, given by $\gamma_j^A=(d_j+d_j^\dagger)/\sqrt{2}$, $\gamma_j^B=i(d_j^\dagger -d_j)/\sqrt{2}$. The Hamiltonian terms of the Kitaev chain in this basis are illustrated in Fig.~\ref{figure: isolated - Majorana}. We name it ``chiral'' because in this basis the chiral symmetry has an especially simple representation, $\mathcal{C} = \sigma_z \otimes \mathbbm{1}_{N\times N}$. The physical quantities in this representation are denoted by the subscript $c$, e.g. $\mathcal{H}_c$, $I_c$. We use it only for the spectrum calculations in appendix~\ref{appendix: excited states}, to highlight how the $A$ components of an eigenstate are mapped by inversion onto its $B$ components.

 \item {\em Site-ordered particle-hole basis} is just a rearranged default basis, with
\begin{equation}
\label{appendix equation: site-wise particle-hole basis}
\hat{\tilde{\Psi}} = (d_1,d_1^\dag,...,d_N,d_N^\dag)^T. 
\end{equation}
We denote the physical quantities in this basis by a $\tilde{}$, e.g. $\tilde{G}^s$, $\tilde{\Sigma}_\alpha$. The transformation between this and the default basis is given by $\hat{\tilde{\Psi}} = U\hat{\Psi}$, with
\begin{equation}
\label{appendix equation: tilded to default basis transformation}
U_{nm} = \left\{ \begin{array}{ll}
                \delta_{m,(n+1)/2} & \textnormal{for } n \textnormal{ odd}\\[2mm]
                \delta_{m,N+n/2} & \textnormal{for } n \textnormal{ even ,}
                 \end{array} \right.  
\end{equation}
with $n,m = 1,...,2N$. Since $U$ is just a permutation matrix, an observable $A$ transforms as $\tilde{A} = U A U^T$. 
This is our intermediate basis in Appendix~\ref{Appendix: Derivation of the current formula}, in which the site-specific Green's functions are expressed most conveniently. 

\item {\em Site-ordered Majorana basis} is a rearranged chiral basis, with
\begin{equation}
\label{appendix equation: site-wise Majorana basis}
\hat{\Psi}_M = (\gamma_1^A,\gamma_1^B,...,\gamma_N^A,\gamma_N^B). 
\end{equation}
The physical quantities in this basis are denoted by the subscript $_M$. The unitary transformation to the default basis, such that $\hat{\Psi}_\mathrm{M}=\mathrm{T}\,\hat{\Psi}$, is given by the matrix $\mathrm{T}$
\begin{align*}
	\mathrm{T}\,=\,\frac{1}{\sqrt{2}}\,\left[\left.\begin{matrix}
	1 & 0\\
	-i &0\\
	0& 1 &0\\
	0&-i&0\\
	&&  \ddots & \ddots\\
	&&& 1 &0\\
	&&& -i &0\\
	&&& 0 & 1\\
	&&& 0 &-i
	\end{matrix}
	\right\vert
	\begin{matrix}
	1 & 0\\
	i &0\\
	0& 1 &0\\
	0&i&0\\
	&&  \ddots & \ddots\\
	&&& 1 &0\\
	&&& i &0\\
	&&& 0 & 1\\
	&&& 0 &i
	\end{matrix}
		\right],
\end{align*}
where "$\vert$" separates the first $N$ and the last $N$ columns. The physical quantities transform as $A_M = \mathrm{T} A \mathrm{T}^\dag$. We use this basis to set up the polynomial sequences in Appendix~\ref{appendix: Tetranacci polynomials}, which in turn define the eigenvectors of the system in Appendix~\ref{appendix: excited states}.
In the Appendix~\ref{appendix: exact expression GF for current} we take advantage of the block-tridiagonal form of the Hamiltonian in this basis (the elements of the Hamiltonian are illustrated in Fig.~\ref{figure: isolated - Majorana}).
\end{enumerate}

\section{Tetranacci polynomials and their closed formula}
\label{appendix: Tetranacci polynomials}
The eigenvectors of Hamiltonians describing the nearest-neighbour hopping in an SSH chain could be obtained using the Fibonacci polynomials\cite{Leumer-2020}, where each term in the polynomial sequence is determined by its immediate neighbours. The presence of further non-diagonal terms in the Hamiltonian will result in more complex polynomial sequences, including more elements in the recursion formula. In the general Kitaev chain, which is equivalent to two SSH-like chains additionally coupled by $\mu$, we must use polynomial sequences where each term is determined by {\em four} preceding ones. These {\em Tetranacci polynomials} are essential for the calculations of the excited state wave functions in Appendix \ref{appendix: excited states} and for the exact calculation of the Green's functions in Appendix \ref{appendix: exact expression GF for current}. 
\subsection{Definition and basic properties}
The eigenvectors of a Kitaev chain can be expressed through two sequences of Tetranacci polynomials, one for the $A$ and one for the $B$ elements of the eigenvectors (later referred to as $\vec{v}_A$ and $\vec{v}_B$). We describe this procedure in Appendix~\ref{appendix: excited states}.\\
The characteristic polynomial of the isolated Kitaev chain, which we will need for the calculation of the Green's functions, can be expressed through {\em four} Tetranacci sequences $x_j,y_j,\chi_j$ and $\ycal_j$, which obey two equivalent sets of coupled equations. The first set is
\begin{subequations}
	\label{appendix equation: coupled recursion formula x_j 1}
\begin{align}
	x_{j+1}&=\frac{-i\mu}{b}\,x_j\,+\,\frac{a}{b}\,x_{j-1}\,+\,\frac{E}{b}\,y_j,\\
	\chi_{j+1}&=\frac{-i\mu}{b}\,\chi_j\,+\,\frac{a}{b}\,\chi_{j-1}\,+\,\frac{E}{b}\,\ycal_j,\\
	y_{j+1}&=\frac{i\mu}{a}\,y_j\,+\,\frac{b}{a}\,y_{j-1}\,+\,\frac{E}{a}\,x_j,\\
	\ycal_{j+1}&=\frac{i\mu}{a}\,\ycal_j\,+\,\frac{b}{a}\,\ycal_{j-1}\,+\,\frac{E}{a}\,\chi_j,
	\label{appendix equation: recursion for ycal 1}
\end{align}
\end{subequations}
and the second set reads
\begin{subequations}
	\label{appendix equation: coupled recursion formula x_j 2}
\begin{align}
	x_{j+1}&=\frac{-i\mu}{b}\,x_j\,+\,\frac{a}{b}\,x_{j-1}\,+\,\frac{E}{a}\,\chi_j,\\
	\chi_{j+1}&=\frac{i\mu}{a}\,\chi_j\,+\,\frac{b}{a}\,\chi_{j-1}\,+\,\frac{E}{b}\,x_j,\\
	y_{j+1}&=\frac{-i\mu}{b}\,y_j\,+\,\frac{a}{b}\,y_{j-1}\,+\,\frac{E}{a}\,\ycal_j,\\
	\ycal_{j+1}&=\frac{i\mu}{a}\,\ycal_j\,+\,\frac{b}{a}\,\ycal_{j-1}\,+\,\frac{E}{b}\,y_j.\label{appendix equation: recursion for ycal 2}
\end{align}
\end{subequations}
Using the relationships defined by these equations, we can decouple the four sequences and find that each of the $x_j$,  $y_j$, $\chi_j$ , $\ycal_j$ polynomials obeys the same recursion formula as $x_j$ (and as the eigenvector entries $\vec{v}_A,\vec{v}_B$),
\begin{align}\label{appendix equation: Tetranacci recursion formula}
	x_{j+2}\,&=\,\frac{E^2+a^2+b^2-\mu^2}{ab}\,x_j\,-\,x_{j-2}\notag\\
	&\quad +\,i\mu \frac{b-a}{ab}\left(x_{j-1}+x_{j+1}\right).
\end{align}
The four sequences differ only in their initial values, given in table \ref{table: initial values of the tetranacci of the pure Kitaev chain}. They appear at first as separated objects, but they are in fact connected by a symmetry relation. By exchanging all $a$ terms by $b$ terms and vice versa and turning the sign of $\mu$ into $-\mu$, $x_j$ ($\chi_j$) transforms into $y_j$ ($\ycal_j$). 

Although the Eqs. \eqref{appendix equation: coupled recursion formula x_j 1} and \eqref{appendix equation: coupled recursion formula x_j 2} are equivalent to Eq. \eqref{appendix equation: Tetranacci recursion formula}, each description has its own advantages, often unseen in the other. For example, the comparison of Eq. \eqref{appendix equation: recursion for ycal 1} with Eq. \eqref{appendix equation: recursion for ycal 2} yields, 
\begin{align}
	b\,\chi_j\,=\,a\,y_j,
\end{align}
and more such relationships can be found. 

Note also that Eq.~\eqref{appendix equation: Tetranacci recursion formula} is invariant under the inversion symmetry and exchange of $\Delta\rightarrow -\Delta$. Further, all four polynomials carry no physical unit and $x_j$, $\ycal_j$ ($\chi_j$, $y_j$) are real (pure imaginary) objects.

\begin{table}[ht]
	\centering
	\caption{The first values of the Tetranacci polynomials $x_j$, $y_j$, $\chi_j$ and $\ycal_j$. }
	\begin{longtable}{p{1.5 cm} p{1.5 cm} p{1.5 cm} p{1.5 cm} p{0.8 cm} }
	\hline\hline \\
	$j$ & $x_j$ &  $\ycal_j$ & $\chi_j$  & $y_j$\\
	\\
	\hline
	\\
	-3  	&  $\frac{i\mu\,b}{a^2}$ & $\frac{-i\mu\,a}{b^2}$ & $\frac{-E}{b}$ & $\frac{-E}{a}$\\
	\\
	-2 		& $\frac{b}{a}$ 	& $\frac{a}{b}$ 		& $0$ 		& $0$\\
	\\
	-1 		& $0$				& $0$					& $0$		& $0$\\ 
	\\
	0		& $1$				& $1$					& $0$		& $0$\\
	\\
	1 		& $\frac{-i\mu}{b}$	     & $\frac{i\mu}{a}$	    & $\frac{E}{b}$ & $\frac{E}{a}$	 \\\\
	\hline\hline 
	\end{longtable}
	\label{table: initial values of the tetranacci of the pure Kitaev chain}
\end{table}

However, the most important reason to present Eqs. \eqref{appendix equation: coupled recursion formula x_j 1} and \eqref{appendix equation: coupled recursion formula x_j 2} is the limiting case of $E=0$, which we need to obtain the conductance formula at zero bias and temperature later. While in Eq. \eqref{appendix equation: Tetranacci recursion formula} seemingly not much happens at $E=0$, in \eqref{appendix equation: coupled recursion formula x_j 1} and \eqref{appendix equation: coupled recursion formula x_j 2} we find that  $x_j$,  $y_j$, $\chi_j$ , $\ycal_j$ become decoupled and obey simplified recursion formulas. We denote the polynomials in the case of $E=0$ with  $x_{j,0}$,  $y_{j,0}$, $\chi_{j,0}$, $\ycal_{j,0}$. The polynomials $\chi_{j,0}\,=\,y_{j,0}\,\equiv\,0$  due to the initial values in table \ref{table: initial values of the tetranacci of the pure Kitaev chain}, and $x_{j,0}$ and $\ycal_{j,0}$ reduce to Fibonacci polynomials \cite{Hoggatt, Webb, Oezvatan-2017} 
\begin{align}\label{appendix equation: case E=0 for x}
	b\,x_{j+1,0}\,&=\,-i\mu\, x_{j,0}\,+\,a\,x_{j-1,0},\\
	\label{appendix equation: case E=0 for y}
	a\,\ycal_{j+1,0}\,&=\,i\mu\, \ycal_{j,0}\,+\,b\,\ycal_{j-1,0}.
\end{align}
A power law ansatz for $x_{j,0}\propto R^j$ ($\ycal_{j,0}\propto \tilde{R}^j$) leads first to the values of $R_\pm$ ($\tilde{R}_\pm$)
\begin{align}\label{appendix equation: definition R pm}
	R_\pm=\frac{-i\mu\pm\sqrt{4\,ab-\mu^2}}{2b},\quad \tilde{R}_\pm=\frac{i\mu\pm\sqrt{4\,ab-\mu^2}}{2a}.
\end{align}
A superposition of $R_\pm^j$ ($\tilde{R}_\pm^j$) leads to
\begin{align}\label{appendix equation: closed form of xj for E=0}
	x_{j,0}\,=\,\frac{R_+^{j+1}-R_-^{j+1}}{R_+ -R_-},\\
	\ycal_{j,0}\,=\,\frac{\tilde{R}_+^{j+1}-\tilde{R}_-^{j+1}}{\tilde{R}_+ -\tilde{R}_-}.\notag
\end{align}
This closed formula for a Fibonacci polynomial is the so-called {\em Binet form}.\cite{Hoggatt, Webb, Oezvatan-2017} The similarity between $R_\pm$ and $\tilde{R}_\pm$ allows us to determine $\ycal_{j,0}$ in terms of $x_{j,0}$
\begin{align}\label{appendix equation: relation between ycal and x at E=0}
	\ycal_{j,0}\,=\,\left(-\frac{b}{a}\right)^j\,x_{j,0},
\end{align}
which leads to many simplifications for the conductance formula later. 

Please notice that the case of $E=0$ in Eq. \eqref{appendix equation: Tetranacci recursion formula} leads always to Fibonacci polynomials even in problems distinct from the Kitaev chain, where the 
Eqs. \eqref{appendix equation: coupled recursion formula x_j 1} and \eqref{appendix equation: coupled recursion formula x_j 2} are unknown. 

A second limiting case exists for $\mu=0$. We see directly from Eq. \eqref{appendix equation: Tetranacci recursion formula}, but not from the Eqs. \eqref{appendix equation: coupled recursion formula x_j 1} and \eqref{appendix equation: coupled recursion formula x_j 2}, that $x_j$,  $y_j$, $\chi_j$ , $\ycal_j$ show again a Fibonacci character, only of a different kind compared with the $E=0$ case. Define $u_j\defl x_{2j}$ ($v_j\defl x_{2j-1}$) and thus $u_j$ ($v_j$) obeys
\begin{align}\label{appendix equation: recursion formula for u,v}
	u_{j+1}\,=\,\frac{E^2+a^2+b^2}{ab}\, u_{j}-u_{j-1},
\end{align}
which mimics the form of Eq. \eqref{appendix equation: case E=0 for x} with different coefficients and a power law ansatz gives their closed form\cite{Leumer-2020}.

From the physical point of view, the two sequences of polynomials $u_{j+1}$ and $v_{j+1}$ construct the Green's functions of the $\mu=0$ case, in which the Kitaev chain can be considered as two decoupled SSH-like chains\cite{Leumer-2020}.
\subsection{The closed form of Tetranacci polynomials and their Fibonacci decomposition}
\label{The closed form of Tetranacci polynomials and their Fibonacci decomposition}
We turn now to the closed formula for any Tetranacci polynomial. In the context of the eigenvectors of the isolated Kitaev chain we will be interested in sequences $x_j$,  $y_j$, $\chi_j$, $\ycal_j$. In the context of Green's functions of a Kitaev chain between two leads the relevant Tetranacci polynomials are $d^x_j$,  $d^y_j$, $d^\chi_j$ , $d^\ycal_j$ discussed further in Appendix \ref{appendix: exact expression GF for current}. 
We shall therefore derive a closed form for $\xi_j$, a general sequence of Tetranacci polynomials obeying Eq. \eqref{appendix equation: Tetranacci recursion formula}, with \emph{arbitrary} initial values $\xi_{-2}$, $\xi_{-1}$, $\xi_{0}$ and $\xi_{1}$. The expressions for $x_j$,  $y_j$, $\chi_j$, $\ycal_j$, $d^x_j$,  $d^y_j$, $d^\chi_j$ and $d^\ycal_j$ can then be obtained by inserting appropriate initial values into the formula for $\xi_j$. 

The idea is to use a power law ansatz $\xi_j \propto r^j$ ($r\neq 0$) as we did in the limiting cases $E=0$ and $\mu=0$ before. We are left to find all zeros of
\begin{align}\label{appendix equation: characteristic equation for tetranacci polynomials}
	r^4\,-\,\zeta\,r^2\,+\,1\,-\,\eta \,(r\,+\,r^3)\,=\,0,
\end{align} 
where we used a shorthand notation for the coefficients in Eq. \eqref{appendix equation: Tetranacci recursion formula}
\begin{align}\label{appendix equation: definition of zeta}
	\zeta &\defl\frac{E^2+a^2+b^2-\mu^2}{ab},\\
	\eta &\defl i\mu\frac{b-a}{ab}.
\end{align}
One can solve for the zeros by dividing Eq. \eqref{appendix equation: characteristic equation for tetranacci polynomials} by $r^2$ and calling $S=r+1/r$. Thus, we have
\begin{align*}
	S^2\,-\,2\,-\,\zeta\,-\,\eta\, S=0,
\end{align*}
and the solutions for $S$ read
\begin{align}\label{appendix equation: form of s12}
	S_{1,2}\,=\,\frac{\eta\pm \sqrt{\eta^2+4(\zeta+2)}}{2}.
\end{align}
Finally, we can get the zeros from $S_{1,2}$. They read 
\begin{align}\label{appendix equation: the four fundamental solutions}
	r_{\pm i}\,=\,\frac{S_i\pm\sqrt{S_i^2-4}}{2},\quad i=1,2.
\end{align}
An expression for the $S_{1,2}$ in terms of two wave numbers $\kappa_1,\kappa_2$,
\begin{equation}
S_{1,2} = 2\cos(\kappa_{1,2}),
\end{equation}
yields directly the physical interpretation of $r_{\pm i}^j$ as plane waves, $r_{\pm i}^j = \exp(\pm i\kappa_i j)$. Since $S_{1,2}$ contains the energy $E$ (via $\zeta$, see Eq. \eqref{appendix equation: definition of zeta}), it connects the energy $E$ and wave numbers $\kappa_{1,2}\in \mathbbm{C}$. Indeed, $S_{1,2} = 2\cos(\kappa_{1,2})$ is the shortest form of the dispersion relation of the Kitaev chain in Eq. \eqref{equation: bulk dispersion relation} and implies directly $E(\kappa_1)=E(\kappa_2)$. Note that $\kappa_{1,2}$ are not quantized so far.
The details of the connection between $r_{\pm i}$, $S_{i}$ and the dispersion relation of the isolated Kitaev chain, the wave vectors and their quantisation rule, is given in Ref. \onlinecite{Leumer-2020}. 

The ansatz for $\xi_j$ is simply
\begin{align}\label{appendix equation: ansatz for arbitrary tetranacci}
	\xi_j\,=\,c_1\,r_{+1}^j+c_2\,r_{-1}^j\,+\,c_3\,r_{+2}^j+c_4\,r_{-2}^j,
\end{align} 
and the coefficients $c_1,\,\ldots,\, c_4$ are fixed by $\xi_{-2},\,\ldots,\,\xi_{1}$. Once the $c_1,\,\ldots,\, c_4$ are known in terms of $\xi_{-2},\,\ldots,\,\xi_{1}$, one can reorder Eq. \eqref{appendix equation: ansatz for arbitrary tetranacci} according to the independent contributions of the initial values. This results in 
\begin{align}\label{appendix equation: tetranacci closed form}
	\xi_j\,=\,\sum\limits_{i=-2}^1 \xi_{i}\,X_i(j),
\end{align}
where the functions $X_i(j)$ depend only on various powers of $r_{\pm 1}$, $r_{\pm 2}$, see Eq. \eqref{appendix equation: gamma -2} - \eqref{appendix equation: gamma 1} below, but not on the values of $\xi_{-2},\,\ldots,\,\xi_{1}$. Hence, changing the values of $\xi_{-2},\,\ldots,\,\xi_{1}$ does not change the functions $X_i(j)$. As one sees directly from Eq. \eqref{appendix equation: tetranacci closed form}, there are constraints on $X_i(j)$, namely
\begin{align}\label{appendix equation: initial values for gamma}
	X_i(j)\,=\,\delta_{i,j}, \quad \mathrm{for~} i,j=\,-2,\,\ldots,\,1,
\end{align}
to ensure that the initial values are assumed by $\xi_j$. One can understand Eq. \eqref{appendix equation: tetranacci closed form} as the counterpart to the Binet form, which is used to determine the closed form expression of Fibonacci polynomials\cite{Webb,Hoggatt, Oezvatan-2017} (see e.g. Eq~\eqref{appendix equation: closed form of xj for E=0}).

Despite the short form of $\xi_j$ in Eq. \eqref{appendix equation: tetranacci closed form}, the formulas for $X_i(j)$ tend to be lengthy, such that we first introduce a shorthand notation for their main pieces. We define the functions $F_{1,2}(j)$ as
\begin{align}
	\label{appendix equation: definition F_1}
	F_1(j)\defl\frac{r^j_{+1}-r^j_{-1}}{r_{+1}-r_{-1}}\,=\,\frac{r^j_{+1}-r^{-j}_{+1}}{r_{+1}-r^{-1}_{+1}},\\
	\label{appendix equation: definition F_2}
	F_2(j)\defl\frac{r^j_{+2}-r^j_{-2}}{r_{+2}-r_{-2}}\,=\,\frac{r^j_{+2}-r^{-j}_{+2}}{r_{+2}-r^{-1}_{+2}},
\end{align}
where the r.h.s of both equalities arise due to $r_{i}\,r_{-i}=1$ for $i=1,2$. Please notice that already $F_{1,2}(j)$ are special solutions of Eq. \eqref{appendix equation: Tetranacci recursion formula}, since they are constructed in terms of the solutions $r_{\pm i}$ (see Eq.~\eqref{appendix equation: F in terms of plane waves}). 

The polynomials $X_i(j)$ read
\begin{align}
	\label{appendix equation: gamma -2}
	X_{-2}(j)\,&=\,\frac{F_2 (j)-F_1 (j)}{S_1-S_2},\\
	X_{-1}(j)\,&=\,\sum_{\sigma=1}^2\frac{F_\sigma (j+2)+ F_\sigma (j-1) F_{\bar{\sigma}}(2) -F_\sigma(3)F_{\bar{\sigma}}(j) }{\left(S_1-S_2\right)^2},\\
	X_{0}(j)&=\,\sum_{\sigma=1}^2\frac{F_\sigma(j+1) F_{\bar{\sigma}}(3)-F_\sigma(j+2) F_{\bar{\sigma}}(2)}{\left(S_1-S_2\right)^2}\notag\\
	&\quad -\sum_{\sigma=1}^2 \frac{F_\sigma(j-1)}{\left(S_1-S_2\right)^2},\\
	\label{appendix equation: gamma 1}
	X_{1}(j)\,&=\,\sum_{\sigma=1}^2\frac{F_\sigma(j+2)+F_\sigma(j)-F_\sigma(j+1)  F_{\bar{\sigma}}(2)}{\left(S_1-S_2\right)^2},
\end{align}
where $\bar{\sigma}$ is meant as "not $\sigma$", e.g. if $\sigma=1$ then we have $\bar{\sigma}=2$ and vice versa. As one sees, the functions $X_i(j)$
are a superposition of the solutions $F_{1,2}(j-x)$ \makebox{($x=-2,-1,0,1$)}, where the coefficients are sometimes $F_{1,2}(2)$ or $F_{1,2}(3)$. Thus, the $X_i(j)$ are Tetranacci polynomials as well. As we saw in Eq. \eqref{appendix equation: ansatz for arbitrary tetranacci}, four initial values are required to fix a solution of Eq. \eqref{appendix equation: Tetranacci recursion formula} and these are given with the selective property in Eq. \eqref{appendix equation: initial values for gamma} for the $X_i(j)$'s. We call the $X_i(j)$ basic or primitive Tetranacci polynomials.

A second proof that the $X_i(j)$ obey the recursion formula in \eqref{appendix equation: Tetranacci recursion formula}, follows directly from Eq. \eqref{appendix equation: tetranacci closed form}. Choosing only one initial value different from zero, e.g. $\xi_j=\delta_{j1}$ for $j=-2,\,\ldots,\,1$, results in
\begin{align*}
	\xi_j\,=\,X_1(j).
\end{align*} 
Similar choices reveal that $\xi_j$ can be equal to only one of the $X_i(j)$. Thus, the $X_i(j)$ must be Tetranacci polynomials. 

The easier form of $x_{j,0}$ in Eq. \eqref{appendix equation: closed form of xj for E=0} cannot be seen from here, since the $r_{\pm i}$ does not reduce to the $R_{\pm}$ at $E=0$. The reason is, that the recursion formulas for $E=0$ and $E\neq 0$ do not transform directly into each other. In the limiting case of $\mu=0$, we find from Eq. \eqref{appendix equation: form of s12} that
\begin{align*}
	S_1\vert_{\mu=0}\,=\,-S_2\vert_{\mu=0},
\end{align*}
yielding
\begin{align*}
	r_{+1}\vert_{\mu=0}\,=\,-r_{-2}\vert_{\mu=0}.
\end{align*}
The effect on $F_{1,2}$ in Eqs. \eqref{appendix equation: definition F_1}, \eqref{appendix equation: definition F_2} is
\begin{align*}
	F_1(j)\vert_{\mu=0}\,=\,(-1)^{j-1}\,F_2(j)\vert_{\mu=0},
\end{align*}
and we find further that
\begin{align*}
	X_{-2}(2l+1)\vert_{\mu=0} \,&=\,0,\\
	X_{0}(2l+1)\vert_{\mu=0} \,&=\,0,\\
	X_{-1}(2l)\vert_{\mu=0} \,&=\,0,\\
	X_{1}(2l)\vert_{\mu=0} \,&=\,0,
\end{align*}
for all values of $l$. Thus, the form of the recursion formula at $\mu=0$ in Eq. \eqref{appendix equation: Tetranacci recursion formula} is respected and the Tetranacci polynomials $\xi_j$ reduce back to Fibonacci polynomials for $\mu=0$. 
This behavior of $\xi_j$ can be also understood in a different way. The definition of the Tetranacci polynomial $F_{1,2}$ is actually a Binet form of Fibonacci polynomials \cite{Hoggatt, Webb, Oezvatan-2017}. One can easily prove that $F_{1,2}$ obey 
\begin{align}\label{appendix equation: Fibonacci recursion formula for F}
	F_i(j+2)\,=\,S_i\,F_i(j+1)-F_i(j),
\end{align} 
for all $j$, with $F_{1,2}(0)=0$, $F_{1,2}(1)=1$. Thus, the closed form of $\xi_j$ can be seen as a superposition of two distinct sequences of Fibonacci polynomials $F_{1,2}$. 

However, one has to account for all four different fundamental solutions $r_{\pm i} (i=1,2)$ in the case of Tetranacci polynomials. Similar to the denominator of a Binet form, which contains the difference of the two fundamental solutions of the corresponding Fibonacci sequence ($ r_{+i}-r_{-i}$ for $F_i(j)$), we find that $S_{1,2}$ adopt this role in the case of Tetranacci polynomials. Note that $S_{1}-S_{2} = r_{+1}+r_{-1}-r_{+2}-r_{-2}$.

\section{Eigenvectors and degeneracies in the spectrum}
\label{appendix: excited states}

\subsection{The general eigenvector problem}
\label{appendix: excited states - general}
We briefly recapitulate here the eigenvector problem investigated in Ref.~\onlinecite{Leumer-2020} and introduce the inversion symmetry before we turn to the degenerate energy eigenvalues. The Kitaev chain Hamiltonian can be expressed in the chiral basis (cf. Eq.~\eqref{appendix equation: chiral basis}) through
\begin{align}\label{equation: Kitaev Hamiltonian in chiral basis}
	\mathcal{H}_c\,=\,\left[\begin{matrix}
	0_{N\times N} & h\\
	h^\dagger & 0_{N\times N}
\end{matrix}			\right],
\end{align}
with $\hat{H}_{\mathrm{KC}}=\frac{1}{2} \hat{\Psi}_c^\dagger \, \mathcal{H}_c\, \hat{\Psi}_c$ and $h_{n,m} = -i \mu \,\delta_{n\,m}+ a \,\delta_{n\,m+1} -  b\,\delta_{n+1\,m}$  for $n,\,m=1,\,\ldots,\,N$, $a = i(\Delta- t)$, $b = i(\Delta+ t)$. For an eigenvector $\vec{w} = \left(\vec{v}_A,\,\vec{v}_B\right)^\mathrm{T}$ of the Hamiltonian $\mathcal{H}_c$ the sublattice vectors $\vec{v}_A\defl \left(\xi_1\,\ldots,\,\xi_N\right)^\mathrm{T}$, $\vec{v}_B\defl \left(\sigma_1\,\ldots,\,\sigma_N\right)^\mathrm{T}$ have to obey
\begin{align}
	\label{appendix equation: obtaining the sublattice vector for A}
	h \,\vec{v}_B &= E\, \vec{v}_A,\, \\
	\label{appendix equation: obtaining the sublattice vector for B}
	h^\dagger \,\vec{v}_A &= E \,\vec{v}_B.
\end{align}
In particular, we consider here exclusively the case of $E\neq 0$, where one can choose all $\xi_n$ ($\sigma_n$) as real (pure imaginary) numbers. Solving for $\vec{v}_A$ grants
\begin{align}
\label{appendix equation: decoupled equation for vA}
	h	h^\dagger\,  \vec{v}_A = E^2 \,\vec{v}_A,
\end{align}
and $\vec{v}_B$ follows then from Eq. \eqref{appendix equation: obtaining the sublattice vector for B}. Importantly, Eq.~\ref{appendix equation: decoupled equation for vA} directly implies that the entries of $\vec{v}_A$ obey the Tetranacci recursion formula
\begin{align}\label{appendix equation: tetranacci recursion formula subblattice vector v}
	\xi_{j+2}\,&=\,\frac{E^2+a^2+b^2-\mu^2}{ab}\,\xi_j\,-\,\xi_{j-2}\notag\\
	&\quad +\,i\mu \frac{b-a}{ab}\left(\xi_{j-1}-\xi_{j+1}\right).
\end{align}
Extending the sequence of $\xi$'s via Eq. \eqref{appendix equation: tetranacci recursion formula subblattice vector v} beyond the range $j=1,...,N$ allows the simplification of the boundary condition to
\begin{align}\label{appendix equation: boundary condition for sublattice vector v}
	\xi_0 = \xi_{N+1} = b\, \xi_1 -a\, \xi_{-1} = b\, \xi_{N+2}- a\, \xi_{N}=0,
\end{align}
while $\vec{v}_A$ still contains only $\xi_1,\,\ldots,\,\xi_N$ and the boundary condition yields after some algebra the quantisation rule\cite{Leumer-2020} given by Eq. \eqref{equation: quantisation rule}. The values of $\kappa_{1,2}$ and $E$ are thus fixed. A sequence obeying Eq. \eqref{appendix equation: tetranacci recursion formula subblattice vector v} requires four initial values, for example $\xi_{-2},\,\xi_{-1},\,\xi_{0},\,\xi_1$, and one can derive the following closed formula
\begin{align}\label{closed formula, entries of eigenvectors}
	\xi_j\,=\,\sum\limits_{i=-2}^1 \xi_{i}\,X_i(j),
\end{align}
where the functions $X_i(j)$, (cf. Eqs. \eqref{appendix equation: gamma -2} - \eqref{appendix equation: gamma 1} in the appendix \ref{The closed form of Tetranacci polynomials and their Fibonacci decomposition}) depend on $E^2$, $t$, $\Delta$, $\mu$ and $N$; they inherit the selective property $X_i(j) = \delta_{i,j}$ for \emph{only} $i,j = -2\,,\ldots,\,1$. As we see from the boundary conditions in Eq. \eqref{appendix equation: boundary condition for sublattice vector v}, $\xi_0=0$ and thus is fixed, while $\xi_{-1} = b \,\xi_1 /a$. In the case of \emph{no degeneracy} we have one degree of freedom and we can choose $\xi_1$ arbitrarily. Consequently, $\xi_{-2}$ is the last missing initial value and can be fixed via $\xi_{N+1}=0$ yielding 
\begin{align}\label{appendix equation: xi minus 2 as function of xi 1}
	\xi_{-2}\,=\,-\xi_1\,\frac{a\,X_1(N+1)+b \,X_{-1}(N+1)}{a\,X_{-2}(N+1)},
\end{align}
in the absence of degeneracy. The eigenvector problem is now solved, since the constraint $b\, \xi_{N+2}- a\, \xi_{N}=0$ quantizes the wave vectors generating the eigenvalue $E$ and thus the values of the functions $X_i(j)$ are known. 

\subsection{Inversion symmetry}
\label{appendix: excited states - inversion                                              }
   
In our previous work we used a rather lengthy method to determine the entries of $\vec{v}_B$ from the ones of $\vec{v}_A$. The inversion symmetry $I$ allows us to pursue a much simpler method as we explain in the following. The modified inversion symmetry of the Kitaev chain which we discussed in Section \ref{section: isolated Kitaev chain}, i.e. the invariance of $\hat{H}_{\mathrm{KC}}$ under the exchange $d_{j}\rightarrow i d_{N+1-j}$, $d^{\dag}_{j}\rightarrow -i d^{\dag}_{N+1-j}$, involves an additional global phase of $i$ for the Nambu spinor. The consequences of the simple inversion symmetry of the original chain ($I: d_j^{(\dagger)} \rightarrow d_{N+1-j}^{(\dagger)}$) can however be explored in a more elegant way, which does not need to introduce an additional phase. In the BdG basis, $I \,\hat{H}_\mathrm{KC}\, I^{-1} =\hat{H}_\mathrm{KC}\vert_{-\Delta}$.  
Written in the basis of $\mathcal{H}_c$, the representation of $I$ is
\begin{align*}
	I_c \,=\,\left[\begin{matrix}
	I_0& \\
	& I_0
	\end{matrix}\right], \quad I_0 = \left[\begin{matrix}
	&& 1\\
	&\rddots \\	
	1\\ 
	\end{matrix}\right],
\end{align*}
where $I_0$ represents the usual inversion operation, i.e. reversing the site order. The use of $I_c$ on the eigenvector problem yields 
\begin{align*}
	\left(h\vert_{-\Delta}\right) \,I_0\vec{v}_B= E\, I_0\vec{v}_A, \quad \left(h^\dagger\vert_{-\Delta}\right)\, I_0\vec{v}_A = E \,I_0\vec{v}_B.
\end{align*}
Importantly, we have that $h\vert_{-\Delta} = - h^\dagger$ and vice versa, transforming the equations for $I_0\, \vec{v}_A$ ($I_0\,\vec{v}_B$) in the ones of $\vec{v}_B$ ($\vec{v}_A$) at $-E$. Recalling that all $\vec{v}_A$ ($\vec{v}_B$) are real (pure imaginary) vectors, we can cancel this sign of $E$ by the now obvious relation between $\vec{v}_A$ and $\vec{v}_B$
\begin{align}
	\vec{v}_A &= \pm i I_0 \vec{v}_B,\\
	\vec{v}_B &= \mp i I_0 \vec{v}_A.
\end{align}
Thus, the entries of $\vec{v}_B$ obey now the simple relation: $\sigma_{N+1-j} = \mp i\, \xi_j$ and a normalized eigenvector $\vec{w}$ is achieved by normalizing $\vec{v}_A$ and division by $\sqrt{2}$. 

In the special case of $E=0$, the degenerate eigenstates are still related by inversion symmetry, but the decoupling of $\vec{v}_A$ and $\vec{v}_B$ allows always to set one of them to zero whereby a relation between $\vec{v}_A$ and $\vec{v}_B$ of the same eigenvector can become invalid.

Once $\vec{v}_A$ is known, one can rewrite the solution in the basis of the fermionic operators $d_j^{(\dagger)}$. After the transformation the electron $d_j$ (hole $d_j^\dag$) part of the quasiparticle state is $\vec{v}_A-i \vec{v}_B$ ($\vec{v}_A+i \vec{v}_B$). The different combination signals opposite behavior under inversion symmetry as captured in Fig. \ref{figure: full spectrum inversion}. Further, after an application of the particle-hole symmetry to the eigenstates, the character of the electron and hole parts under inversion symmetry changes into the opposite, since the exchange $E\rightarrow -E$ means $\vec{v}_B \rightarrow -\vec{v}_B$ while keeping the same $v_A$.  
\subsection{Degenerate energy levels}
\label{Appendix: energy crossings}
The important starting point for the case of degeneracies is Fig. \ref{figure: full spectrum particle-hole}, where we see that for \emph{specific} values of $t$, $\Delta$ and $\mu$ a crossing in the spectrum occurs, which naturally depends also on $N$. We exclude here from consideration the cases of $E=0$, $t^2= \Delta^2$ ($ab= 0$) and $t\, \Delta = 0$, because they are already known. Further, we consider $t$, $\Delta$ as fixed while $\mu$ can be varied to achieve a degeneracy. 

Let us begin by inspecting the degree of the degeneracy, and assume initially that we have $D\ge 2$ degenerate eigenvectors $\vec{v}^{\,(d)} = \left(\vec{v}_A^{\,(d)},\,\vec{v}_B^{\,(d)}\right)^\mathrm{T}$ with $d=1,\,\ldots,\,D$ and all $\vec{v}^{\,(d)} = \left(\vec{v}_A^{\,(d)},\,\vec{v}_B^{\,(d)}\right)^\mathrm{T}$ have to obey the Eqs. \eqref{appendix equation: obtaining the sublattice vector for A}- \eqref{closed formula, entries of eigenvectors}. We continue with almost the same notation as above, where we change only $\xi_j$ ($\sigma_j$) into $\xi_j^{\,(d)}$ ($\sigma_j^{\,(d)}$) for clarity. The eigenstates are still determined by the quantization rule in Eq. \eqref{equation: quantisation rule}, and in the following we will obtain the required further constraint on Eq. \eqref{equation: quantisation rule} needed for the eigenstates to be degenerate. 

The case of degenerate eigenvectors has to be treated carefully, since their superposition can break the connection between $\vec{v}_A^{\,(d)}$ and $\vec{v}_B^{\,(d)}$ via inversion symmetry. Nonetheless, once the value of the energy is known all information of $\vec{v}^{\,(d)}$ is still contained in $\vec{v}_A^{\,(d)}$, since $\vec{v}_B^{\,(d)}$ follows from $h^\dagger \,\vec{v}_A^{\,(d)} = E \,\vec{v}_B^{\,(d)}$. Furthermore, for given values of $t$, $\Delta$ and $\mu$, the functions $X_i(j)$ in Eq. \eqref{closed formula, entries of eigenvectors} differ only for states with different energy, therefore the $\vec{v}_A^{\,(d)}$ are defined only by distinct initial values $\xi_{-2}^{\,(d)},\,\ldots,\,\xi_1^{\,(d)}$. Thus one can build and exploit special superpositions of those eigenstates yielding 
\begin{align}\label{appendix equation: initial values for degenerated eigenstates1}
	\xi_1^{(1)}&=1,\quad \xi_{-2}^{(1)}=0,\\
	\label{appendix equation: initial values for degenerated eigenstates2}
	\xi_1^{(2)}&=0,\quad \xi_{-2}^{(2)}=1.
\end{align} 
The boundary condition in Eq. \eqref{appendix equation: boundary condition for sublattice vector v} demands $\xi^{(0)} = 0$, $\xi^{(d)}_{-1}=b \,\xi^{(d)}_{1} /a$ and thus fixes $\vec{v}_A^{\,(d)}$. Note that the Eqs. \eqref{appendix equation: initial values for degenerated eigenstates1}, \eqref{appendix equation: initial values for degenerated eigenstates2} imply $D=2$, i.e. only twofold degeneracies are allowed, since beyond $\xi_1$ and $\xi_{-2}$ there are no further degrees of freedom to exploit. As we see next, Eq. \eqref{appendix equation: xi minus 2 as function of xi 1} which formerly coupled $\xi_1^{(d)}$ and $\xi_{-2}^{(d)}$ becomes indeed invalid for the new superpositions. Returning to the boundary condition in Eq. \eqref{appendix equation: boundary condition for sublattice vector v}, we get further constraints, namely
\begin{align}
	\label{appendix equation: x minus two N+1}
	X_{-2}(N+1)&=0,\\
	\label{appendix equation: x minus two N+2, N}
	b\,X_{-2}(N+2)-a  \,X_{-2}(N)&=0,\\
	\label{appendix equation: x1 two N+1}
	a\,X_{1}(N+1)\,+\,b X_{-1}(N+1)&=0,\\[2mm]
	b\left[a X_{1}(N+2)+ b  X_{-1}(N+2)\right]\notag\\
	\label{appendix equation: x1 two N+2, 2}
		-a\left[a  X_{1}(N)+  b  X_{-1}(N)\right]&=0,
\end{align}
implying a division of zero by zero in Eq. \eqref{appendix equation: xi minus 2 as function of xi 1}. Further, the Eqs. \eqref{appendix equation: x minus two N+1} - \eqref{appendix equation: x1 two N+2, 2} show that the boundary condition splits into two parts for $N+1$ and for $N+2$, $N$ which is the constraint on Eq. \eqref{equation: quantisation rule} for which we have been looking. 

As we have discussed in Appendix \ref{The closed form of Tetranacci polynomials and their Fibonacci decomposition},
the four functions $X_i(j)$ are constructed with the help of two special functions $F_{1,2}$, see Eq. \eqref{appendix equation: gamma -2} - \eqref{appendix equation: gamma 1}, which are are nothing else than standing waves,
\begin{align}
\label{appendix equation: F in terms of plane waves}
	F_{1,2} (j) \,=\,\frac{\sin\left(\kappa_{1,2}\, j\right)}{\sin\left(\kappa_{1,2}\right)}
\end{align} 
at site $j$, constructed from the plane waves $r_{+1,+2} = e^{i \kappa_{1,2}}$ as follows from Eqs. \eqref{appendix equation: the four fundamental solutions}, \eqref{appendix equation: definition F_1}, \eqref{appendix equation: definition F_2}.

Now, we can solve for $\kappa_{1,2}$. We get first from Eq.\eqref{appendix equation: x minus two N+1} two constraints: $S_1-S_2\neq0$, i.e. $\kappa_{1}\neq \pm \kappa_2$, and $F_{1} (N+1)=F_{2} (N+1)$. Second, these two restrictions used on Eq. \eqref{appendix equation: x1 two N+1} together with exploiting the properties of $F_{1,2}$ (for example Eq. \eqref{appendix equation: Fibonacci recursion formula for F}), give us a familiar expression, namely
\begin{align}
	a\,X_{-2}(N+2)-b  \,X_{-2}(N)=0,
\end{align} 
which is almost Eq. \eqref{appendix equation: x minus two N+2, N}. Thus, $X_{-2}(N+2)= X_{-2}(N)=0$ holds, or equivalently $F_{1} (N+2)=F_{2} (N+2)$ and $F_{1} (N)=F_{2} (N)$. This imposes a further constraint on $\kappa_{1,2}$ to obey $F_{1,2}(N+1)=0$. Thus $\kappa_{1,2}= n \pi /(N+1)$, $n= 1,\ldots,\,N$. 

The combinations of different values of $\kappa_{1,2}$ yield both the positions of strict and of avoided crossings in the $(\mu,E)$ plane. The values of $\mu$ follow from Eq. \eqref{equation: kappa 1 2 depend on mu} after converting the values of $\kappa_{1,2}$ into $\kappa_{\Sigma, \Delta}$. The energy $E$, in turn, can be obtained from the dispersion relation, either Eq. \eqref{equation: bulk dispersion relation} or Eq. \eqref{equation: bulk dispersion relation, seconda versione}. Whether these $(\mu,E)$ pairs define strict or avoided crossings is determined by the general quantization rule in Eq. \eqref{equation: quantisation rule}, considering the following facts. (i) With $\kappa_{1,2}= n_{1,2} \pi /(N+1)$ the values for $\kappa_{\Sigma,\Delta}$ are either both half-integer or both integer multiples of $\pi/(N+1)$. (ii) The entire derivation for $\kappa_{1,2}$ is invariant under the exchange $\kappa_1\rightarrow \pm \kappa_2$ and $\kappa_1 \rightarrow -\kappa_1$, hence, without loss of generality we can demand $\kappa_1 > \kappa_2$.  (iii) By virtue of Eq.~\eqref{equation: kappa 1 2 depend on mu} $\kappa_1 + \kappa_2 \neq \pi$, except for $N$ odd and $\mu=0$. In the end, we find that only $\kappa_{\Sigma,\Delta}$ which are integer multiples of $\pi/(N+1)$ satisfy the quantisation rule \eqref{equation: quantisation rule} for arbitrary value of $\Delta$. Thus the selection rule for strict crossings can be expressed in terms of $\kappa_{\Sigma,\Delta}$, demanding that $\kappa_\Sigma >\kappa_\Delta$, and resulting in the requirement of $\kappa_{\Sigma,\Delta}$ being integer multiples of $\pi/(N+1)$ as stated in Eqs.~\eqref{equation: crossing, excited states, kappa sigma values},\eqref{equation: crossing, excited states, kappa delta values}. The half-integer multiples satisfy Eq. \eqref{equation: quantisation rule} only if $\Delta=0$, hence in a superconducting chain they always define avoided crossings.

\section{Derivation of the current formula}
\label{Appendix: Derivation of the current formula} 
The electronic current (for fixed spin) in the left lead is,
\begin{align}
I_{L}(t)\,=\,-e \,\langle \dot{N_{L}}\rangle,
\end{align}
where $e$ is the elementary charge and $N_{L} = \sum\limits_k c_{kL}^\dagger c_{kL}$. The specific choice of the tunneling Hamiltonian $H_\mathrm{T}$ in Eq.~\eqref{equation: Tunneling Hamiltonian} leads to the explicit expression 
\begin{align}\label{appendix equation: current formula, in scalar quantities}
I_{L}(t)\,=\,-\frac{ie}{\hbar} \sum\limits_k \left(t_L\, \langle d_1^\dagger (t) \, c_{kL}(t)\rangle \,-t_L^*\,  \langle c_{kL}^\dagger(t)\, d_1(t)				\rangle\right),
\end{align}
where the superconductivity is contained inside the time evolution of the creation and annihilation operators. Further, we shall use a $2\times 2$ matrix notation for the fermionic Green's functions (GF) whose entries are defined via $D_j\defl(d_j,\,d_j^\dagger)^\mathrm{T}$ as
\begin{align}
\left(\mathbf{G}^>_{ij}(t,t')\right)_{n\,m}\,&\defl\, -\frac{i}{\hbar} \langle \left( D_i(t)\right)_n \,\left( D_j(t')\right)^\dagger_m \rangle,\\
\left(\mathbf{G}^<_{ij}(t,t')\right)_{n\,m}\,&\defl\, ~\frac{i}{\hbar} \langle \left( D_j(t')\right)^\dagger_m \,\left( D_i(t)\right)_n \rangle,\\
\left(\mathbf{G}^r_{ij}(t,t')\right)_{n\,m}\,&\defl\, -\frac{i}{\hbar} \theta(t-t') \langle\left\{ \left( D_i(t)\right)_n ,\,\left( D_j(t')\right)^\dagger_m \right\}\rangle,\\
\left(\mathbf{G}^a_{ij}(t,t')\right)_{n\,m}\,&\defl\, \frac{i}{\hbar} \theta(t'-t) \langle \left\{ \left( D_i(t)\right)_n ,\,\left( D_j(t')\right)^\dagger_m \right\}\rangle,
\end{align}
where $\{\cdot,\cdot\}$ denotes the anticommutator and $n,m = 1,2$.
All kinds of Green's functions, such as $\mathbf{G}^s_{k\alpha}(t,t'),\,\mathbf{G}^s_{j k\alpha}(t,t')$ for $s=<,r,a,>$, are defined analogously with $C_{k\alpha}\defl(c_{k\alpha},\,c_{k\alpha}^\dagger)^\mathrm{T}$ instead of $D_{j}$. Please keep in mind that the NEGF formalism uses a large variety of inter-related Green's functions. \\
In the following we will denote all $2\times2$ matrices with a bold font, to keep them distinct from the $2N\times2N$ matrices used everywhere else.

A convenient expression for the current in terms of those $2\times 2$ matrices is
\begin{align}
	I_L(t) = -e\sum\limits_k \,\mathrm{Re}\left\{\mathrm{Tr}\left[\left(\begin{matrix}
	t_L &0\\
	0& t_L^*
	\end{matrix}
	\right)
	\,\mathbf{G}_{kL\,1}^<(t,t)\right]\right\}.
\end{align}
Starting from the equation of motion for the relevant Green's functions, and using standard relations between them together with the Langreth rules~\cite{Flensberg,Jauho,diVentra,Ryndyk,Langreth}, we find the steady state current
\begin{align}\label{appenix equation: current in terms of 2x2 matrices}
	I_L &= -e \int\limits_{\mathbbm{R}}\,\frac{d\omega}{2\pi} \mathrm{Tr}\left\{
	\tau_z\,\left[\,\mathbf{\Sigma}_L^r(\omega)\,\mathbf{G}_{11}^<(\omega)\right.\right.\notag
	\\
	&\qquad \left.\left. +\mathbf{\Sigma}_L^<(\omega)\,\mathbf{G}_{11}^a(\omega)
		\right]\right\},
\end{align}
with 
\begin{align}
	\mathbf{\Sigma}_\alpha^r(\omega)\,&=\,\lim\limits_{\eta \rightarrow 0}\sum\limits_k \vert t_\alpha(k)\vert^2 \,\left[\begin{matrix}
	\frac{1}{\hbar \omega -\epsilon_{k\alpha}+i \eta} & 0\\
	0 &	\frac{1}{\hbar \omega +\epsilon_{k\alpha}+i \eta}
	\end{matrix}\right],\\
	\mathbf{\Sigma}_\alpha^<(\omega)\,&=\,2\pi\,i\,\sum\limits_k \vert t_\alpha(k)\vert^2 \,\left[\begin{matrix}
\delta(\hbar \omega-\epsilon_{k\alpha}) & 0\\
	0 &	\delta(\hbar \omega+\epsilon_{k\alpha})
	\end{matrix}\right]\times\notag\\
&\qquad	\times
	\left[\begin{matrix}
f(\hbar \omega -eV_\alpha) & 0\\
	0 &	f(\hbar \omega +eV_\alpha)
	\end{matrix}\right].
\end{align}
The lesser Green's function matrices $\mathbf{G}_{11}^{<,a}$ involve only the first site of the Kitaev chain and carry information about the coupling of this site with both the rest of the chain and the leads. To obtain them it is convenient to work in the site-ordered particle-hole basis (cf. Eq.~\eqref{appendix equation: site-wise particle-hole basis}), where the $2\times2$ matrices introduced above become the building blocks of $\tilde{G}^s$ ($s=<,r,a,>$),
\begin{align}\label{appendix equation: definintion of tilde G^s}
	\tilde{G}^s\,=\,\left[\begin{matrix}
	\mathbf{G}_{11}^{s} & \ldots &\mathbf{G}_{1N}^{s}\\
	\vdots & & \vdots\\
	\mathbf{G}_{N1}^{s} & \ldots &\mathbf{G}_{NN}^{s}
	\end{matrix}
	\right].
\end{align}
We find that $\tilde{G}^{r}$ obeys
\begin{align}\label{appendix equation: g retarted tilde as the inverse of}
	\left[\left(\hbar\omega+ i \eta\right)\mathbbm{1}_{2N}-\tilde{\mathcal{H}}-\tilde{\Sigma}_L^r-\tilde{\Sigma}_R^r
		\right]\tilde{G}^{r}=\mathbbm{1}_{2N},
\end{align}
with the self-energy matrices $\tilde{\Sigma}_{L,R}^s$ ($s=<,r,a,>$) given by
\begin{align}
	\tilde{\Sigma}_L^s\,=\,\left[\begin{matrix}
	\mathbf{\Sigma}_L^s & \mathbf{0} & \ldots &\mathbf{0}\\
	\mathbf{0} &	\mathbf{0} & \ldots &  \mathbf{0} \\
	\vdots &  \vdots &\ddots  & \vdots\\
	\mathbf{0} & \mathbf{0} & \ldots &\mathbf{0}
	\end{matrix}
	\right]_{2N  \times 2N},\\
	\tilde{\Sigma}_R^s\,=\,\left[\begin{matrix}
 \mathbf{0}& \ldots & \mathbf{0}&\mathbf{0}\\
 \vdots&	\ddots &\vdots & \vdots  \\
	\mathbf{0}&  \ldots &\mathbf{0}  & \mathbf{0}\\
	\mathbf{0} & \ldots &\mathbf{0} &	\mathbf{\Sigma}_R^s
	\end{matrix}
	\right]_{2N  \times 2N},
\end{align}
where $\mathbf{0}$ is the $2 \times 2$ matrix filled with zeros. The Hamiltonian $\tilde{\mathcal{H}}$ reads
\begin{align}
	\tilde{\mathcal{H}}\,=\,\left[\begin{matrix}
	-\mu \,\tau_z & \boldsymbol{\alpha}  & &\\
	\boldsymbol{\alpha}^\dagger & -\mu \,\tau_z & \boldsymbol{\alpha} \\
	& \ddots& \ddots& \ddots\\
	& &	\boldsymbol{\alpha}^\dagger & -\mu \,\tau_z & \boldsymbol{\alpha}\\
	& & & 	\boldsymbol{\alpha}^\dagger & -\mu \,\tau_z
	\end{matrix}
	\right]_{2N  \times 2N},
\end{align}
where we kept the Pauli matrix $\tau_z$ in regular font, and the matrix $\boldsymbol{\alpha}$  
\begin{align}\label{appendix equation: matrix alpha}
	 \boldsymbol{\alpha}\,=\,\left[\begin{matrix}
	 -t & -\Delta\\
	 \Delta & t
	 \end{matrix}
	 \right]
\end{align}
accounts for nearest neighbor terms. Further, $\tilde{G}^<$ obeys
\begin{align}\label{appendix equation: keldysh equation in so basis for 2Nx2N matrices}
	\tilde{G}^<\,=\,\tilde{G}^r \left(\tilde{\Sigma}^<_{L}+\tilde{\Sigma}^<_{R}\right)\tilde{G}^a,
\end{align}
with $\tilde{G}^a = \left(\tilde{G}^r\right)^\dagger$ so that all ingredients of Eq. \eqref{appenix equation: current in terms of 2x2 matrices} are in principle known. The trace, and the sparsity of the self-energies $\tilde{\Sigma}^s_{L,R}$ allow us to express the current
\begin{align}
	I_L &= -e \int\limits_{\mathbbm{R}}\,\frac{d\omega}{2\pi} 
\mathrm{Tr}\left\{\mathbbm{1}_N\otimes\tau_z
	\,\left[\tilde{\Sigma}_L^r(\omega)\,\tilde{G}^<(\omega)\right.\right.\notag
	\\
	&\qquad \left.\left. +\tilde{\Sigma}_L^<(\omega)\,\tilde{G}^a(\omega)
	\right]\right\},
\end{align}
in terms of these $2N \times 2N$ matrices. We define  $\tilde{\Gamma}_\alpha \defl -2 \,\mathrm{Im}(\tilde{\Sigma}_\alpha^r)= i(\tilde{\Sigma}_\alpha^r-\tilde{\Sigma}_\alpha^a)$ and with
\begin{align}
	\tilde{F}_\alpha\,=\,\mathbbm{1}_N\otimes\,\left[\begin{matrix}
	f(\hbar \omega -eV_\alpha) & 0\\
	0 &	f(\hbar \omega +eV_\alpha)
	\end{matrix}\right],
\end{align}
it follows that $\tilde{\Sigma}_\alpha^< =\tilde{\Gamma}_\alpha \,	\tilde{F}_\alpha$. Since the current is a real quantity, i.e. $2I_L = I_L + I_L^\dagger$, we find the appealing form\cite{Meir-Wingreen-1992}
\begin{align}
	I_L &=i \frac{e}{2} \int\limits_{\mathbbm{R}}\,\frac{d\omega}{2\pi} 
	\mathrm{Tr}\left\{\left(\mathbbm{1}_N\otimes\tau_z\right)
	\,\tilde{\Gamma}_L\left[\,\tilde{G}^<+\tilde{F}_L \left(\tilde{G}^r-\tilde{G}^a\right)
	\right]\right\}.
\end{align}
Expressing all quantities in the default basis \eqref{appendix equation: default basis} yields directly Eq. \eqref{equation: current formula bdg basis}. The corresponding expressions of the self-energies and Green's functions are explicitly given in appendix \ref{appendix: B, GF and self energies in the BDG basis}. 

Finding the analytical form of the conductance demands first a simplification towards Eq. \eqref{equation: current formula, two processes}, which mostly consists of taking the trace and using the sparsity of the self-energies. This procedure is performed at best by using Eq. \eqref{appenix equation: current in terms of 2x2 matrices} and a basis transformation \eqref{appendix equation: tilded to default basis transformation} at the end. We find from Eq. \eqref{appendix equation: keldysh equation in so basis for 2Nx2N matrices} that
\begin{align}
	\mathbf{G}_{11}^< = \mathbf{G}^r_{11}~\mathbf{\Sigma}_{L}^<~\mathbf{G}^a_{11}+
	\mathbf{G}^r_{1N}~\mathbf{\Sigma}_{R}^<~\mathbf{G}^a_{N1}
\end{align}
and Eq. \eqref{appendix equation: g retarted tilde as the inverse of} yields first $\tilde{G}^r-\tilde{G}^a = -i \,\tilde{G}^r (\tilde{\Gamma}_L+ \tilde{\Gamma}_R) \tilde{G}^a$ and thus 
\begin{align}
	\mathbf{G}^r_{11}-\mathbf{G}^a_{11} = -i\left[\mathbf{G}^r_{11}~\mathbf{\Gamma}_{L}~\mathbf{G}^a_{11}+\mathbf{G}^r_{1N}~\mathbf{\Gamma}_{R}~\mathbf{G}^a_{N1}\right].
\end{align}
With $2I_L = I_L + I_L^\dagger$ it follows from Eq. \eqref{appenix equation: current in terms of 2x2 matrices} that
\begin{align}\label{appendix equation: current formula last step before the trace is taken}
	I_L &= i\frac{e}{2} \int\limits_{\mathbbm{R}} \frac{d\omega}{2\pi} \mathrm{Tr}\left\{
	\tau_z\,\left[
		\mathbf{\Gamma}_L~\mathbf{G}_{11}^r~\mathbf{\Sigma}_L^<~\mathbf{G}_{11}^a \right.\right.
		\notag\\
		 & \qquad\qquad -\,	\mathbf{\Sigma}_L^<~
	 \mathbf{G}_{11}^r~\mathbf{\Gamma}_L~\mathbf{G}_{11}^a \notag\\
	& \qquad \qquad +\,	\mathbf{\Gamma}_L~\mathbf{G}_{1N}^r~\mathbf{\Sigma}_R^<~\mathbf{G}_{N1}^a
	\notag\\
	&\qquad \left.\left. \qquad -\,	\mathbf{\Sigma}_L^<~
	\mathbf{G}_{1N}^r~\mathbf{\Gamma}_R~\mathbf{G}_{N1}^a 	\right]\right\} ,
\end{align}
where the $2\times 2$ broadening matrices read 
\begin{align}
	\mathbf{\Gamma}_{\alpha} = \left[\begin{matrix}
	\Gamma_\alpha^- & 0\\
	0 & \Gamma_\alpha^+
	\end{matrix}\right]
\end{align}
with the abbreviations $\Gamma_\alpha^\pm =2\pi \sum\limits_{k} \vert t_\alpha (k)\vert^2\,\delta\left(\hbar \omega \pm\epsilon_{k\alpha}\right)$. In order to shorten the expression of the trace, we define
\begin{align}
	f_\alpha^\pm \defl f(\hbar \omega \pm eV_\alpha)
\end{align}
and after a bit of algebra one finds  
\begin{align}
	&i\,\mathrm{Tr}\left\{
	\tau_z\,\left[
	\mathbf{\Gamma}_L~\mathbf{G}_{11}^r~\mathbf{\Sigma}_L^<~\mathbf{G}_{11}^a 
	-\,	\mathbf{\Sigma}_L^<~
	\mathbf{G}_{11}^r~\mathbf{\Gamma}_L~\mathbf{G}_{11}^a\right]\right\}\notag\\
	\label{appendix equation: Andreev term in site ordered basis}
	&=\,\Gamma_{L}^-\,\Gamma_{L}^+ \left(\vert \tilde{G}^r_{1,2}(\omega)\vert^2 + \vert \tilde{G}^r_{2,1}(\omega)\vert^2
	\right)\,\left[f_L^- -f_L^+
	\right], \\
	\label{equation: non-local Gs}
	&i\,\mathrm{Tr}\left\{
	\tau_z\,\left[
	\mathbf{\Gamma}_L~\mathbf{G}_{1N}^r~\mathbf{\Sigma}_R^<~\mathbf{G}_{N1}^a 
	-\,	\mathbf{\Sigma}_L^<~
	\mathbf{G}_{1N}^r~\mathbf{\Gamma}_R~\mathbf{G}_{N1}^a\right]\right\}\notag\\
	&=\,\Gamma_{L}^-\,\Gamma_{R}^- \,\vert \tilde{G}^r_{1,2N-1}(\omega)\vert^2\,\left[f_L^--f_R^-
	\right] \notag\\
	&\quad +\,\Gamma_{L}^+\,\Gamma_{R}^+ \,\vert \tilde{G}^r_{2,2N}(\omega)\vert^2\,\left[f_R^+ -f_L^+
	\right]\notag\\
	&\quad +\,\Gamma_{L}^-\,\Gamma_{R}^+ \,\vert \tilde{G}^r_{1,2N}(\omega)\vert^2\,\left[f_L^- -f_R^+
	\right]\notag\\
	&\quad +\,\Gamma_{L}^+\,\Gamma_{R}^- \,\vert \tilde{G}^r_{2,2N-1}(\omega)\vert^2\,\left[f_R^- -f_L^+
	\right].
\end{align}
In contrast to Eq. \eqref{equation: current formula, two processes}, where only electronic contributions are used, in Eqs.~\eqref{appendix equation: Andreev term in site ordered basis}, \eqref{equation: non-local Gs} we have six terms for both electronic and hole degrees of freedom, and a factor of $1/2$ in front of Eq. \eqref{appendix equation: current formula last step before the trace is taken} to avoid overcounting. The following steps will further reduce the number of terms. 

Throughout our approach, we considered $t$ and $\Delta$ as real quantities. Hence, $\tilde{\mathcal{H}}$ is a symmetric matrix. Since $\tilde{\Sigma}_\alpha^r$ are symmetric too, we have that $\tilde{G}^r_{i,j} =\tilde{G}^r_{j,i}$. This yields in Eq. \eqref{appendix equation: Andreev term in site ordered basis} a factor of $2$.

Further, the particle-hole symmetry gives
\begin{align}\label{appendix equation: particle hole symmetry}
	\left(\mathbbm{1}_N \otimes \sigma_x\right) \,\left[\tilde{G}^r(-\omega)\right]^*\,	\left( \mathbbm{1}_N\otimes \sigma_x\right)\,=\,-\tilde{G}^r(\omega),
\end{align}
where "$*$" denotes the complex conjugation. The use of Eq. \eqref{appendix equation: particle hole symmetry} on $\tilde{G}^r(\omega)$ and observing its particular action on the entries of the $2\times 2$ block $\mathbf{G}^r_{1N}$ yields
\begin{align}\label{appendix equation: symmetry on G retared tile, omega and -omega 1}
	\tilde{G}^r_{2,2N}(\omega) &= -\left[\tilde{G}^r_{1,2N-1}(-\omega)\right]^*,\\
	\label{appendix equation: symmetry on G retared tile, omega and -omega 2}
	\tilde{G}^r_{1,2N}(\omega) &= -\left[\tilde{G}^r_{2,2N-1}(-\omega)\right]^* .
\end{align}
Since $\Gamma^\pm_\alpha(\omega) = \Gamma^\mp_\alpha(-\omega)$ holds, one has simply to split the integration in Eq. \eqref{appendix equation: current formula last step before the trace is taken} into two parts. After a substitution of $\omega \rightarrow -\omega$ and the use of the relations in  Eqs. \eqref{appendix equation: symmetry on G retared tile, omega and -omega 1}, \eqref{appendix equation: symmetry on G retared tile, omega and -omega 2}, we find that
\begin{align}\label{appendix equation: semi final formula for the current}
	I_L \,&=\,e\int\limits_\mathbbm{R}\frac{d\omega}{2\pi}\,\left\{
	\Gamma_{L}^-(\omega)\Gamma_{L}^+(\omega) \,\vert \tilde{G}^r_{1,2}(\omega) \vert^2
	\left[f_L^--f_L^+\right]\right.\notag\\
	&\quad  +\,\Gamma_{L}^-(\omega)\Gamma_{R}^-(\omega) \,\vert \tilde{G}^r_{1,2N-1}(\omega) \vert^2
	\left[f_L^--f_R^-\right]\notag	\\
	&\quad \left. +\,\Gamma_{L}^-(\omega)\Gamma_{R}^+(\omega) \,\vert \tilde{G}^r_{1,2N}(\omega) \vert^2
	\left[f_L^--f_R^+\right]	\right\}
\end{align}
which is already very close to Eq.\eqref{equation: current formula, two processes}; we need now a basis transformation given by Eq.~\eqref{appendix equation: tilded to default basis transformation}. The necessary entries of $\tilde{G}^r$ transform as
\begin{align*}
	\tilde{G}^r_{1,2}&=  G^r_{1,N+1},\\
	\tilde{G}^r_{1,2N-1}&=  G^r_{1,N},\\
	\tilde{G}^r_{1,2N}&=  G^r_{1,2N},
\end{align*}
and inserting this in Eq. \eqref{appendix equation: semi final formula for the current} with the substitution $E=\hbar\omega$ leads almost directly to Eq. \eqref{equation: current formula, two processes}, though the bias still remains to be set.\\
The use of the mean field technique breaks the conservation of the number of particles, if \emph{fixed} values of $\Delta$ are used and thus $I_L\neq -I_R$. For correctness one has to the use self-consistently calculated profile of $\Delta$, since that replaces correctly two operators with their mean values and the number of particles is (implicitly) conserved. On the other side, one obviously prefers to avoid the self-consistency cycle. After we obtain $I_R$, we find that $I_L$=$-I_R$ holds for $\Gamma_L=\Gamma_R$ and symmetrically applied bias ($\eta=1/2$, i.e. $V_L=V/2$, $V_R=-V/2$), without demanding the self-consistently calculated $\Delta$.~\cite{Lim-2012,LevyYeyati-1995} This trick sets the internal supercurrent to zero and allows the use of fixed values of $\Delta$. As a second effect the crossed Andreev term $G^r_{1,2N}$ does not contribute to the current, since the difference of the Fermi functions $f_L^--f_R^+$ is always zero for $\eta=1/2$. 

\section{Matrix expressions in the (standard) Bogoliubov de Gennes basis}
\label{appendix: B, GF and self energies in the BDG basis}
The use of the default basis $\hat{\Psi}=(d_1,\,\ldots, d_N,\,d_1^\dagger,\,\ldots, d_N^\dagger)^\mathrm{T}$ gives an intuitive understanding of the current formula, since the entries of the Hamiltonian, the self-energies and the Green's functions are ordered first in the particle/hole subspace and second in the real space position. For example, $G^r_{1,N}$ describes the transport of an electron from site $j=1$ to site $j=N$, where it leaves the Kitaev chain as an electron to the right lead. We present here the matrices used in Eq. \eqref{equation: current formula bdg basis}. The BdG Hamiltonian $\mathcal{H}$ reads 
\begin{align}\label{equation: BdG Hamiltonian in real space in its full glory}
	\mathcal{H}\,=\,\left[\begin{matrix}
	C & S\\
	S^\dagger & -C
	\end{matrix}\right]_{\small 2N \times 2N},
\end{align}
with $\hat{H}_\mathrm{KC}\,=\,\frac{1}{2}\hat{\Psi}^\dagger	\mathcal{H}\hat{\Psi}$,  $\hat{H}_\mathrm{KC}$ being given by  Eq. \eqref{equation: Kit. Hamiltonian, fermionic operators, realspace}. The matrices $C$ and $S$ are
\begin{align}\label{equation: Matrix C}
	C\,=\,\left[\begin{matrix}
	-\mu & -t  		\\
	-t&-\mu & -t 	\\
	& -t &-\mu & -t &\\
	&  & \ddots & \ddots &\ddots &\\
	&&& -t &-\mu & -t\\
	&&&& -t &-\mu & -t\\
	&&&&& -t &-\mu
	\end{matrix}			\right]_{\small N \times N},
	\\\label{equation: Matrix S}
	S\,=\,\left[\begin{matrix}
	0 & \Delta  		\\
	-\Delta&0 & \Delta 	\\
	& -\Delta &0 & \Delta &\\
	&  & \ddots & \ddots &\ddots &\\
	&&& -\Delta &0 & \Delta\\
	&&&& -\Delta &0 & \Delta\\
	&&&&& -\Delta &0
	\end{matrix}\right]_{\small N \times N}.
\end{align}
Due to the choice of the tunneling Hamiltonian $H_L$ in Eq. \eqref{equation: Tunneling Hamiltonian}, the self-energies $\Sigma_L^r$ and $\Sigma_R^r$ are sparse matrices ($i,j=1,\,\ldots,\,2N$) 
\begin{align}
	\label{appendix equation: self energy Sigma L}
	\left(\Sigma_L^r\right)_{i,j}\,&=\,\delta_{1i} \delta_{1j}~\Omega_{L-}\,+\,\delta_{\small N+1,i} \delta_{\small N+1,j}~\Omega_{L+},\\
	\label{appendix equation: self energy Sigma R}
	\left(\Sigma_R^r\right)_{i,j}\,&=\,\delta_{Ni} \delta_{Nj}~\Omega_{R-}\,+\,\delta_{\small i,2N} \delta_{\small 2N,j}~\Omega_{R+}
\end{align}
acting only on the first and last site. We used here the abbreviations
\begin{align}\label{appendix equation: entries of the self energies}
	\Omega_{\alpha\pm}\,=\,\lim\limits_{\eta \rightarrow 0}\,\sum\limits_k  \frac{\vert t_\alpha (k)\vert^2}{E+i \eta\pm\epsilon_{k\alpha}},\quad \alpha=L,R,
\end{align}
where the index $-$ ($+$) accounts for particles (holes). In general the finite life time introduced by the self energies is given by the imaginary part $\mathrm{Im}\left(\Omega_{\alpha\pm}\right)=-\pi\sum_k \vert t_\alpha (k)\vert^2 $\mbox{$\delta(E\pm\epsilon_{k\alpha})$}$\defr -\gamma_{\alpha}^\pm$. In the special case of the wide band limit the functions $\Omega_{\alpha\pm}$ don't depend on $E$ and become $\Omega_{\alpha\pm}=-i\gamma_\alpha$ from the main text. \\
Returning to the general case, the matrices $\Gamma_\alpha$ follow from
\begin{align*}
	\Gamma_\alpha(E)=-2\, \mathrm{Im}(\Sigma_\alpha^r), \qquad \alpha = L,R.
\end{align*}
The retarded Green's function $G^r$ is given by 
\begin{align*}
	G^r\,=\,\left[E\mathbbm{1}_{2N}-\mathcal{H}_{\mathrm{BdG}}-\Sigma_L^r-\Sigma_R^r\right]^{-1},
\end{align*}
and the advanced Green's function obeys $G^a(E)=\left[ G^r(E)\right]^\dagger$. The Fermi Dirac distribution $f(E)$ is contained in the matrix $F_\alpha$ such that
\begin{align*}
	F_\alpha\,=\,\left[\begin{matrix}
	\mathbbm{1}_N\,f(E-eV_\alpha) &\\
	& \mathbbm{1}_N\,f(E+eV_\alpha) 
	\end{matrix}
	\right],
\end{align*}
where $V_\alpha$ denotes the shift of the chemical potential at contact $\alpha=L,R$. Finally, the lesser Green's function $G^<(E)$ reads
\begin{align}
	G^<(E)\,=\,i\,G^r \,\left[\sum\limits_{\alpha=L,R}F_\alpha\Gamma_\alpha(E)\right]\,G^a.
\end{align}
\section{The exact form of the Green's functions $G^r_{1,N+1}$, $G^r_{1,N}$, $G^r_{1,2N}$}
\label{appendix: exact expression GF for current}
The entries of the retarded Greens function $G^r_{1,N}$ $G^r_{1,N+1}$ and $G^r_{1,2N}$ in the default basis can be obtained analytically. The calculations are most conveniently performed in the site-ordered Majorana basis defined in Eq.~\eqref{appendix equation: site-wise Majorana basis}, since the Kitaev Hamiltonian and the self energies are reshaped into a block tridiagonal matrix, see Eq. \eqref{appendix equation: Kitaev hamiltonian and self energies in Majorana basis} below. Keeping in mind that $G^s = \mathrm{T}^\dag G_M^s \mathrm{T}$, after a bit of algebra one finds that
\begin{align*}
	G^r_{1,N+1}\,&=\,\frac{1}{2}\left\{ \left(G^r_\mathrm{M}\right)_{11}-\left( G^r_\mathrm{M}\right)_{22} + i 
		\left[\left( G^r_\mathrm{M}\right)_{12}+\left( G^r_\mathrm{M}\right)_{21}\right]	\right\},\\
	G^r_{1,N}\,&=\,\frac{1}{2}\left\{ \left(G^r_\mathrm{M}\right)_{1,2N-1}+\left( G^r_\mathrm{M}\right)_{2,2N}\right. \\
		&\quad + i 	\left.\left[\left( G^r_\mathrm{M}\right)_{2,2N-1}-\left( G^r_\mathrm{M}\right)_{1,2N}\right]	\right\}.\\
	G^r_{1,2N}\,&=\,\frac{1}{2}\left\{ \left(G^r_\mathrm{M}\right)_{1,2N-1}-\left( G^r_\mathrm{M}\right)_{2,2N}\right. \\
		&\quad + i 	\left.\left[\left( G^r_\mathrm{M}\right)_{2,2N-1}+\left( G^r_\mathrm{M}\right)_{1,2N}\right]	\right\}.
\end{align*}
Although several other entries of the inverted matrix are required to obtain the entries $G^r_{1,N+1}$, $G^r_{1,N}$ and $G^r_{1,2N}$ after the transformation, the inversion \emph{can} be performed analytically. As it turns out, see Eq. \eqref{appendix equation: characteristic polynomial} below, the problem involves a non-linear combination of polynomials and the basis transformation allows the decomposition.

In the case of $N\neq 1$, the retarded Green's function $G^r_\mathrm{M}$ is the inverse of
\begin{align}\label{appendix equation: Kitaev hamiltonian and self energies in Majorana basis}
	\mathcal{M}\defl E\,\mathbbm{1}_{2N}-\mathcal{H}_\mathrm{M}-\Sigma^r_{L,\mathrm{M}}-\Sigma^r_{R,\mathrm{M}}\,=\notag\\
	=\left[
	\begin{matrix}
	\mathbf{A}_L & \mathbf{B}\\
	\mathbf{C} & \mathbf{A}_2 & \mathbf{B}\\
	& \mathbf{C} & \mathbf{A}_3 & \mathbf{B}\\
	&	& \ddots & \ddots & \ddots\\
	& & & \mathbf{C} & \mathbf{A}_{N-1} & \mathbf{B}\\
	& &   & &\mathbf{C} & \mathbf{A}_R
	\end{matrix}
	\right],
\end{align}
with 
\begin{align*}
	\mathbf{A}_j\,&=\,\left[
	\begin{matrix}
	E & i \mu\\
	-i \mu & E
\end{matrix}\right],\quad \mathbf{C}^\dagger =  \mathbf{B} =\left[
	\begin{matrix}
	0 & -a\\
	-b  & 0
\end{matrix}\right],\\
	\mathbf{A}_\alpha\,&=\mathbf{A}_2\,+\,\left[
	\begin{matrix}
	\sigma_{\alpha,p} & i\,\sigma_{\alpha,m}\\
	-i\,\sigma_{\alpha,m} & \sigma_{\alpha,p}\\
\end{matrix}\right],
\end{align*}
and $j=2\,,\ldots,N-1$, $a=i(\Delta-t)$, $b=i(t+\Delta)$, $\sigma_{\alpha,p}=-(\Omega_{\alpha+}+\Omega_{\alpha-})/2$, $\sigma_{\alpha,m}=(\Omega_{\alpha+}-\Omega_{\alpha-})/2$. In the case of $N=1$, $G^r_\mathrm{M}$ is the inverse of
\begin{align*}
	\mathbf{A}_2 +\sum\limits_{\alpha=L,R}\left[
	\begin{matrix}
	\sigma_{\alpha,p} & i\,\sigma_{\alpha,m}\\
	-i\,\sigma_{\alpha,m} & \sigma_{\alpha,p}\\
\end{matrix}\right].
\end{align*}
The final results for $G^r_{1,N+1}$, $G^r_{1,N}$ and $G^r_{1,2N}$ unite the cases $N=1$, $N\neq 1$ and so we drop this distinction. The required entries of $G^r_\mathrm{M}$ are obtained via the adjoint matrix technique, where one needs to calculate only determinants. The matrix form in Eq. \eqref{appendix equation: Kitaev hamiltonian and self energies in Majorana basis} allows us to use the method explained in Ref. [\onlinecite{Molinari}], which entails the inversion of the $\mathbf{B}$-type matrices. The calculation of $\mathrm{det}\left(\mathcal{M}\right)$ is straightforward, but notice that the seemingly unimportant structure of the $\mathbf{B}$ matrices is the key here.
The matrices $\mathbf{B}$, $\mathbf{B}^{-1}$ are off-diagonal, which yields simpler coefficients in the recursion formula and is the reason to use a basis of Majorana operators. A site-ordered fermionic basis (cf. Eq.~\eqref{appendix equation: site-wise particle-hole basis}), replaces $\mathbf{B}$ with $-\boldsymbol{\alpha}$ from Eq. \eqref{appendix equation: matrix alpha} and the calculation cannot be performed that easily.

Nevertheless, obtaining the entries of the adjoint matrix themselves requires even further tricks, which we cannot cover here. To give only one example: the minors of $\mathcal{M}$ which we have to calculate for the entries of $G^r$ are not of the same block tridiagonal form as $\mathcal{M}$ itself, one column and one row is missing. One has thus to extend the minors of $\mathcal{M}$ to $2N\times 2N$ without changing the value of the determinant, while at the same time restoring the same block tridiagonal shape. For the Andreev contributions one has simply to add a row and a column, which contain only zeros except one single "$1$" at position $(1,1)$ of this new matrix. Laplace's expansion shows that the value of the determinant is unchanged, but the newly formed first upper/ lower off-diagonal block is not invertible. In order to cure this, one has to consider an entire sequence of matrices, which converge back to the former, etc. Furthermore, once this calculation is accomplished, still a different procedure has to adopted to calculate the direct and crossed Andreev terms. 

We shall therefore simply give below the closed formulae for the relevant Green's functions, and justify their form a posteriori. For example, if one calculates first $\mathrm{det}\left(E\mathbbm{1}_{2N}-\mathcal{H}_\mathrm{M}\right)$, which is essentially the characteristic polynomial and straightforward\footnote{In the strict approach, one has to exclude that $ab=0$ in order to arrive at the following results. The case of $ab=0$ follows by taking the limit of $a\rightarrow0$ and/or $b\rightarrow 0$ at the end. The full result is smooth in $a$ and $b$ as one can proof easily. However, in Ref. \onlinecite{Molinari} each inversion of $B$ is 
countered by a multiplication with $\mathrm{det}(B)$ for cancellation, but both operations enter at different levels in the procedure. Hence, $ab\neq0$ is only a technical but not a physical restriction.} to derive with Ref. \onlinecite{Molinari}, one finds that
\begin{align}\label{appendix equation: characteristic polynomial}
	\mathrm{det}\left(E\,\mathbbm{1}_{2N}-\mathcal{H}_\mathrm{M}\right)\,=\,(-ab)^{N}\,\left(x_N\,\ycal_N\,-\,y_N\,\chi_N\right),
\end{align}
where the functions $x_N$, $\ycal_N$, $y_N$, $\chi_N$ are Tetranacci polynomials of order $N$, as discussed in Appendix \ref{appendix: Tetranacci polynomials}. They obey the recursion formula~\cite{Leumer-2020} Eq.~\eqref{appendix equation: Tetranacci recursion formula}, which we repeat here for the sake of convenience,
\begin{align}\label{appendix equation: tetranacci recursion formula for the characteristic polynomial}
	x_{j+2}\,&=\,\frac{E^2+a^2+b^2-\mu^2}{ab}\,x_j\,-\,x_{j-2}\notag\\
	&\quad +\,i\mu \frac{b-a}{ab}\left(x_{j-1}-x_{j+1}\right),
\end{align}
with the initial values for $x_N$, $\ycal_N$, $y_N$, $\chi_N$  given in table \ref{table: initial values of the tetranacci of the pure Kitaev chain}. 
In order to generalize the result in Eq. \eqref{appendix equation: characteristic polynomial} to the case including the self energies, one should remember that the self-energies act only on the first/last site; the interior of the matrix $\mathcal{M}$ in Eq. \eqref{appendix equation: Kitaev hamiltonian and self energies in Majorana basis} is not affected by them. Importantly, the recursion formula in Eq. \eqref{appendix equation: tetranacci recursion formula for the characteristic polynomial} is a consequence of this structure. This justifies an attempt (as it turns out, successful) to solve our problem using Tetranacci polynomials. 

Following the technique of Ref. \onlinecite{Molinari}, one can define the polynomials $d^y_j$, $d^x_j$, $d^\chi_j$ and $d^\ycal_j$ as a superposition of $x_j$, $\ycal_j$, $y_j$, $\chi_j$
\begin{align}\label{appendix equation: definition of tetranaccis in presence of right contact 1}
	d^y_j\,&\defl\,\sigma_{R,p}~x_{j-1}\,+i\,\sigma_{R,m}~y_{j-1}\,+\,a\,y_j,\\
	d^\ycal_j\,&\defl\,\sigma_{R,p}~\chi_{j-1}\,+i\,\sigma_{R,m}~\ycal_{j-1}\,+\,a\,\ycal_j,\\
	d^x_j\,&\defl\,\sigma_{R,p}~y_{j-1}\,-i\,\sigma_{R,m}~x_{j-1}\,+\,b\,x_j,\\
	\label{appendix equation: definition of tetranaccis in presence of right contact 4}
	d^\chi_j\,&\defl\,\sigma_{R,p}~\ycal_{j-1}\,-i\,\sigma_{R,m}~\chi_{j-1}\,+\,b\,\chi_j,
\end{align}
including the entries of the right self-energy as coefficients. Physical intuition leads us to believe that, similar to $d^y_j,\,d^\ycal_j,\,d^x_j$ and $d^\chi_j$, also Tetranacci polynomials including only the left self-energy exist. Our use of Eqs. \eqref{appendix equation: definition of tetranaccis in presence of right contact 1} - \eqref{appendix equation: definition of tetranaccis in presence of right contact 4} is only a matter of the chosen technique.

In the end, one finds
\begin{align}\label{appendix equation: Determinant of M}
	\frac{\mathrm{det}\left(\mathcal{M}\right)}{(-ab)^{N-1}}\,&=\,d^y_N\,d^\chi_N-d^x_N\, d^\ycal_N\notag\\
		&\quad +\,\frac{\sigma_{L,m}^2-\sigma_{L,p}^2}{ab}\,\left[d^y_{N-1}\,d^\chi_{N-1}\,-\,d^x_{N-1} \,d^\ycal_{N-1}\right]\notag\\
		&\quad +\frac{\sigma_{L,p}}{b} \left[d^y_N \,d^x_{N-1}\,-\,d^x_N \,d^y_{N-1}\right]\notag\\
		&\quad +\frac{\sigma_{L,p}}{a} \left[d^\chi_N \,d^\ycal_{N-1}\,-\,d^\ycal_N \,d^\chi_{N-1}\right]\notag\\
		&\quad + i\frac{\sigma_{L,m}}{a} \left[d^y_N\, d^\chi_{N-1}\,-\,d^x_N \,d^\ycal_{N-1}\right]\notag\\
		&\quad + i\frac{\sigma_{L,m}}{b} \left[d^\ycal_N\, d^x_{N-1}\,-\,d^\chi_N\, d^y_{N-1}\right],
\end{align}

and the entries $G^r_{1,N+1}$, $G^r_{1,N}$ and $G^r_{1,2N}$ read
\begin{align}
	\label{appendix equation: retarded GF Andreev}
	G^r_{1,N+1}\,\frac{2\mathrm{det}\left(\mathcal{M}\right)}{(-ab)^{N-2}}\,&=\,\frac{b^2}{a}\left[d^{\ycal}_{N-2}\,d^\chi_{N-1}\,-\,d^\ycal_{N-1} \,d^						\chi_{N-2}\right]\notag\\
	& \quad +\frac{a^2}{b}\left[d^y_{N-2}\, d^x_{N-1}\,-\,d^y_{N-1}\, d^x_{N-2}\right]\notag\\
	& \quad +i a\left[d^\chi_{N-1}\, d^y_{N-2}\,-\,d^\ycal_{N-1} \,d^x_{N-2}\right]\notag\\
	& \quad -i b\left[d^x_{N-1}\, d^\ycal_{N-2}\,-\,d^y_{N-1}\, d^\chi_{N-2}\right],
\end{align}
\begin{align}
	\label{appendix equation: retarded GF crossed}
	G^r_{1,2N}\frac{2\mathrm{det}\left(\mathcal{M}\right)}{(-ab)^{N-1}}\,&=\,\frac{b}{a}\left[d^\chi_{N-2} \,- \,i \,d^\ycal_{N-2}	\right]\notag\\
	&\quad -\frac{a}{b}\left[d^y_{N-2} \,+ \,i \,d^x_{N-2}\right]\notag\\
	&\quad + \left(E-\Omega_{L,+}\,-\,\mu\right)\times\notag\\
	&\quad \times \left[\frac{d^x_{N-1}-i d^y_{N-1}}{b}-\frac{d^\ycal_{N-1}+i d^\chi_{N-1}}{a}
	\right],
\end{align}
\begin{align}
	\label{appendix equation: retarded GF direct}
	G^r_{1,N}\frac{2\mathrm{det}\left(\mathcal{M}\right)}{(-ab)^{N-1}}\,&=\,\frac{b}{a}\left[d^\chi_{N-2} + i\, d^\ycal_{N-2}	\right]\notag\\
	&\quad +\frac{a}{b}\left[d^y_{N-2}\, -\, i\, d^x_{N-2}\right]\notag\\
	&\quad + \left(E-\Omega_{L,+}\,-\,\mu\right)\times\notag\\
	&\quad \times \left[\frac{d^x_{N-1}+i d^y_{N-1}}{b}+\frac{d^\ycal_{N-1}-i d^\chi_{N-1}}{a}
	\right].
\end{align}
The results for $\mathrm{det}\left(\mathcal{M}\right)$, $G^r_{1,N+1}$, $G^r_{1,N}$ and $G^r_{1,2N}$ hold for \emph{all} values of $N$, $t$, $\Delta$, $\mu$, $E$ and the wide band limit is not used yet. Notice that these functions do not diverge at $t=\pm \Delta$, i.e $a=0$ or $b=0$. The reason is that all denominators contain only $a's$ and $b's$, which are exactly canceled by the prefactors $(-ab)^{N-x}$ for $x=1,2$. This statement is obvious after a look into Eq. \eqref{appendix equation: Kitaev hamiltonian and self energies in Majorana basis}, since no entries of the matrix diverge there. Strictly speaking, one has to take the limit $a$ ($b$) $\rightarrow 0$ and not to evaluate at $a=0$ ($b=0$), but this is merely a numerical issue. 

Taking a closer look to the Andreev contribution in panel a) in Fig. \ref{figure: differential conductance t/d = 4.1}, one observes very light and thin lines representing Andreev conduction minima outside the conduction gap, which intertwine with the darker ones, representing the maxima. They are caused by two features of $G^r_{1,N+1}$: first, it contains Tetranacci polynomials $d_j$ also for $j=N-1,N-2$ and $j=N-3$; second, these polynomials enter here as a product, therefore in a higher order than in $G^r_{1,N}$, and at their zeros the Andreev transmission is suppressed more strongly than the direct transmission.

\section{Conductance formula}
\label{appendix: Conductance formula}
The conductance follows from Eq. \eqref{equation: current formula, two processes} by its derivative w.r.t. the bias in the zero bias limit. In the wide band limit at $T=0\,$K we find
\begin{align}\label{appendix equation: direct conductance}
	G_D\,&=\,4\,\frac{e^2}{h}\,\gamma_L\gamma_R\, \vert G^r_{1,N}\vert^2_{E=0},\\
	\label{appendix equation: Andreev conductance}
	G_A\,&=\,4\,\frac{e^2}{h}\gamma_L^2\,\vert G^r_{1,N+1} \vert^2_{E=0},
\end{align}
and of course $G=G_D+G_A$. The necessary Green's functions are given by the Eqs. \eqref{appendix equation: Determinant of M} - \eqref{appendix equation: retarded GF direct} and we have only to evaluate them at $E=0$. In turn one should focus first on the Tetranacci polynomials $d^y_j$, $	d^\ycal_j$, $d^x_j$, $d^\chi_j$ from Eqs. \eqref{appendix equation: definition of tetranaccis in presence of right contact 1} -\eqref{appendix equation: definition of tetranaccis in presence of right contact 4}. At $E=0$ they reduce to

\begin{align*}
	d^y_j\vert_{E=0}\,&=\,i\gamma_R~x_{j-1,0},\\
	d^\chi_j\vert_{E=0}\,&=\,i\gamma_R~\ycal_{j-1,0},\\
	d^\ycal_j\vert_{E=0}\,&=\,a\,\ycal_{j,0},\\
	d^x_j\vert_{E=0}\,&=\,b\,x_{j,0}.
\end{align*}
We use Eq. \eqref{appendix equation: relation between ycal and x at E=0} to eliminate $\ycal_{j,0}$ and we find in a first step after some algebra that
\begin{align}\label{appendix equation: det of M at E=0 first step}
	\left.\mathrm{det}\left(\mathcal{M}\right)\right\vert_{E=0}\,b^{2-2N}\,&=\,b^2\,x^2_{N,0}-x^2_{N-1,0} \left(\gamma_{L}^2+\gamma_R^2\right),\notag\\
	&\quad +\,x^2_{N-2,0} \frac{\gamma^2_{L}\gamma^2_{R}}{b^2},\notag\\
	&\quad -\,\gamma_L\gamma_{R} \left(x^2_{N-1,0}-x_{N,0} \,x_{N-2,0}\right)	\times\notag\\
	&\qquad \times \frac{a^{2N-2}+b^{2N-2}}{(-ab)^{N-1}}.
\end{align}
The key to shorten the last expression and to further simplifications is the function $g_s$ ($s=\pm 1$)
\begin{align}
	g_s&\defl b^{N-1} \left[is\,b\,x_{N,0}\,+\,i\,x_{N-1,0}\left(\gamma_{R}-s\gamma_{L}\right)\right.\notag\\
	&\qquad \qquad -\frac{i}{b} \left. x_{N-2,0}\, \gamma_L\gamma_R\right],
\end{align}
since one gets that
\begin{align}\label{appendix equation: absolute square of gs}
	-\vert b^{1-N} \,g_s\vert^2\,&=\,b^2\,x_{N,0}^2 -\, x^2_{N-1,0} (\gamma_{R}-s\gamma_{L})^2\notag\\
	&\quad +\, x^2_{N-2,0} \frac{\gamma^2_{L}\gamma^2_{R}}{b^2},\notag\\
	&\quad -\,2s\,\gamma_{L}\gamma_R\, x_{N,0}\, x_{N-2,0},
\end{align}
which is very close to the expression of  $\left.\mathrm{det}\left(\mathcal{M}\right)\right\vert_{E=0}$ in Eq. \eqref{appendix equation: det of M at E=0 first step}. In order to obtain the equality in Eq. \eqref{appendix equation: absolute square of gs} one has to use that $s^2=1$ and that $x_{j,0}$ is a real valued function, see Eq. \eqref{appendix equation: case E=0 for x} and table \ref{table: initial values of the tetranacci of the pure Kitaev chain}. The last identity we need to simplify $	\left.\mathrm{det}\left(\mathcal{M}\right)\right\vert_{E=0}$ reads
\begin{align}\label{appendix equation: special identitiy for fibonacci}
	x_{j-1,0}^2-x_{j,0}\,x_{j-2,0}\,=\,\left(-\frac{a}{b}\right)^{j-1},
\end{align}
which follows directly from Eq. \eqref{appendix equation: closed form of xj for E=0} and the fact that $R_+\,R_- = -a/b$ with $R_\pm$ from Eq. \eqref{appendix equation: definition R pm}. Adding and subtracting the term $2s\,\gamma_L\,\gamma_R \,(x_{N-1,0}^2-x_{N,0}\, x_{N-2,0})$ to  	$\left.\mathrm{det}\left(\mathcal{M}\right)\right\vert_{E=0}$ and using the Eqs. \eqref{appendix equation: absolute square of gs}, \eqref{appendix equation: special identitiy for fibonacci} yields
\begin{align}\label{appendix equation: expression for det(M) in terms of g_s}
	\left.\mathrm{det}\left(\mathcal{M}\right)\right\vert_{E=0}=(-1)^{N} \vert g_s\vert^2 - \gamma_{L}\gamma_{R} \left[a^{N-1}+ s (-b)^{N-1} \right]^2,
\end{align}
where a factor $(-1)^{N-1}$ occurs for taking $b^{N-1}$ out of the absolute value in Eq. \eqref{appendix equation: absolute square of gs}.

Finally, the simplifications of the entries $G^r_{1,N+1}$ and $G^r_{1,N}$ at $E=0$ starting from Eqs. \eqref{appendix equation: retarded GF Andreev}-\eqref{appendix equation: retarded GF direct} read
\begin{align}
	\label{appendix equation: G^r, direct at E=0}
	\left.G^r_{1,N} \right\vert_{E=0}\,&=\,(-1)^{N-1} ~\frac{a^{N-1}\,+\, (-b)^{N-1}}{2\, \mathrm{det}\left.\left(\mathcal{M}\right)\right\vert_{E=0} }~ g_-,\\
	\label{appendix equation: G^r, Andreev at E=0}
	\left.G^r_{1,N+1} \right\vert_{E=0}\,&=\,-i\, \gamma_R ~\frac{a^{2N-2}\,-\, b^{2N-2}}{2\, \mathrm{det}\left.\left(\mathcal{M}\right)\right\vert_{E=0} },\\
	\left.G^r_{1,2N} \right\vert_{E=0}\,&=\,(-1)^{N} ~\frac{a^{N-1}\,-\, (-b)^{N-1}}{2\, \mathrm{det}\left.\left(\mathcal{M}\right)\right\vert_{E=0} }~ g_+,
\end{align}
where we give the result of $\left.G^r_{1,2N} \right\vert_{E=0}$ only for completeness. The use of the Eqs. \eqref{appendix equation: G^r, direct at E=0} - \eqref{appendix equation: G^r, Andreev at E=0} together with Eqs. \eqref{appendix equation: direct conductance} - \eqref{appendix equation: Andreev conductance} yields to the expressions \eqref{equation: direct conductance term} -\eqref{equation: Andreev conductance term}, as we show now.

The function $q_s$ from Eq. \eqref{equation: definition of q_s} is constructed such that $\vert q_s\vert^2 = \vert g_s\vert^2$. Further we have $b = i\,p$, $a = -i\,m$ with $p=t+\Delta$ and $m=t-\Delta$. In a first step we get for the total conductance $G= G_D+G_A$
\begin{widetext}
	\begin{align}
		\frac{\left\vert 2\, \mathrm{det}\left.\left(\mathcal{M}\right)\right\vert_{E=0}\right\vert^2}{4\gamma_L \gamma_R} \,\frac{h}{e^2}\,G\,&=\,
		\left(p^{N-1}+m^{N-1} \right)^2 \vert g_-\vert^2\,+\,\gamma_L\,\gamma_R \left(p^{2N-2}-m^{2N-2} \right)^2 \notag\\
		&\,=	\left(p^{N-1}+m^{N-1} \right)^2 \left[\vert g_-\vert^2 + \gamma_L\gamma_R \left(m^{N-1}-p^{N-1} \right)^2\right],
	\end{align}
\end{widetext}
after reorganizing the terms arising from Eqs. \eqref{appendix equation: G^r, direct at E=0} - \eqref{appendix equation: G^r, Andreev at E=0}. The use of Eq. \eqref{appendix equation: expression for det(M) in terms of g_s} yields
\begin{align*}
	\left\vert \, \mathrm{det}\left.\left(\mathcal{M}\right)\right\vert_{E=0}\right\vert = \left[\vert g_s\vert^2 + \gamma_L\gamma_R \left(m^{N-1}+s\,p^{N-1} \right)^2\right]
\end{align*}
and the total conductance becomes
\begin{align}
	G = \frac{e^2}{h}\,\frac{\gamma_L\,\gamma_R \,\left(p^{N-1}+m^{N-1}\right)^2}{\vert g_+\vert^2 +\gamma_L\,\gamma_R \,\left(p^{N-1}+m^{N-1}\right)^2}.
\end{align}
Since $\vert g_s\vert^2 = \vert q_s\vert^2$ holds we find the conductance according to Eq. \eqref{equation: Conductance, wide band limit zero temperature}.

%

\end{document}